\documentclass[aps,prx,twocolumn,superscriptaddress,showpacs,floatfix,preprintnumbers]{revtex4-2}
\usepackage[utf8]{inputenc}

\usepackage{cancel}

\usepackage[dvipsnames]{xcolor}      
\definecolor{lcolor}{rgb}{0.5,0,0}
\definecolor{citcolor}{rgb}{0,0.0,1}

\usepackage[breaklinks,colorlinks,urlcolor=blue,citecolor=citcolor,linkcolor=lcolor,linktoc=all]{hyperref}
\usepackage{color}
\usepackage{graphicx}	
\graphicspath{{./fig/}}
\usepackage[utf8]{inputenc}
\usepackage{amsmath} 
\usepackage{amssymb}
\usepackage[ragged]{footmisc}
\usepackage{slashed} 
\usepackage{float}
\usepackage[capitalise]{cleveref}
\usepackage{dsfont}

\usepackage[normalem]{ulem} 

\newcommand\sumint[1]{\int\kern-1.5em\sum\nolimits_{#1}}

\makeatletter
\g@addto@macro\bfseries{\boldmath}
\makeatother

\usepackage{tikz}
\usepackage[customcolors]{hf-tikz}
\usepackage{mciteplus}

\usetikzlibrary{arrows,cd,shapes,decorations.pathmorphing,decorations.markings,shadings}
\tikzset{
  big arrow/.style={
    decoration={markings,mark=at position 1 with {\arrow[scale=4,#1]{>}}},
    postaction={decorate},
    shorten >=0.4pt},
  big arrow/.default=blue}

\newcommand{\bbone}{\text{\usefont{U}{bbold}{m}{n}1}}
\MakeRobust{\bbone}
\newcommand{\eq}{Eq.~}

\newcommand{\nr}[1]{(\ref{#1})}

\newcommand{\Lbar}{\overline{\Lambda}}

\newcommand{\pt}{{\mathbf{p}}}
\newcommand{\qt}{{\mathbf{q}}}

\DeclareMathOperator*{\SumInt}{%
\mathchoice%
  {\ooalign{$\displaystyle\sum$\cr\hidewidth$\displaystyle\int$\hidewidth\cr}}
  {\ooalign{\raisebox{.14\height}{\scalebox{.7}{$\textstyle\sum$}}\cr\hidewidth$\textstyle\int$\hidewidth\cr}}
  {\ooalign{\raisebox{.2\height}{\scalebox{.6}{$\scriptstyle\sum$}}\cr$\scriptstyle\int$\cr}}
  {\ooalign{\raisebox{.2\height}{\scalebox{.6}{$\scriptstyle\sum$}}\cr$\scriptstyle\int$\cr}}
}

\DeclareMathOperator*{\SumIntprime}{%
\mathchoice%
  {\ooalign{$\displaystyle\sum$\cr\hidewidth$\displaystyle\;\int^{\prime}$\hidewidth\cr}}
  {\ooalign{\hspace{-3cm}\raisebox{.14\height}{\scalebox{.7}{$\textstyle\sum$}}\cr\hidewidth$\textstyle\int$\hidewidth\cr}}
  {\ooalign{\hspace{-3cm}\raisebox{.2\height}{\scalebox{.6}{$\scriptstyle\sum$}}\cr$\scriptstyle\int$\cr}}
  {\ooalign{\hspace{-3cm}\raisebox{.2\height}{\scalebox{.6}{$\scriptstyle\sum$}}\cr$\scriptstyle\int$\cr}}
}

\DeclareFontFamily{U}{wncy}{}
\DeclareFontShape{U}{wncy}{m}{n}{<->wncyr10}{}
\DeclareSymbolFont{mcy}{U}{wncy}{m}{n}
\DeclareMathSymbol{\Sh}{\mathord}{mcy}{"58}

\begin{document}

\title{Cosmological phase transitions without high-temperature expansions}

\preprint{HIP-2025-20/TH}
\author{Pablo Navarrete}
\email{pablo.navarrete@helsinki.fi}
\affiliation{Department of Physics and Helsinki Institute of Physics, P.O.~Box 64, FI-00014 University of Helsinki, Finland}
\author{Risto Paatelainen}
\email{risto.paatelainen@helsinki.fi}
\affiliation{Department of Physics and Helsinki Institute of Physics, P.O.~Box 64, FI-00014 University of Helsinki, Finland}
\author{Kaapo Seppänen}
\email{kaapo.seppanen@helsinki.fi}
\affiliation{Department of Physics and Helsinki Institute of Physics, P.O.~Box 64, FI-00014 University of Helsinki, Finland}
\author{Tuomas~V.~I.~Tenkanen}
\email{tuomas.tenkanen@helsinki.fi}
\affiliation{Department of Physics and Helsinki Institute of Physics, P.O.~Box 64, FI-00014 University of Helsinki, Finland}

\begin{abstract}
We introduce a new framework for perturbatively computing equilibrium thermodynamic properties of cosmological phase transitions to high loop orders, using the full four‑dimensional resummed thermal effective potential and avoiding the limitations of standard high‑temperature approximations. By systematically disentangling the physics of hard and soft momentum scales, our approach unifies their treatment within a single expression, enabling consistent handling of both vacuum and thermal divergences across all mass regimes. This core innovation enables the efficient numerical evaluation of massive multiloop thermal sum-integrals, achieved through a finite-temperature generalization of Loop-Tree Duality -- an advanced algorithmic technique originally developed to render vacuum Feynman integrals numerically tractable via Monte Carlo methods. As a proof of principle, we apply the framework to a scalar-Yukawa model, presenting a complete two-loop calculation and a novel three-loop extension -- the first fully massive three-loop sum-integral computation without relying on high-temperature expansions. Our approach opens the door to precise perturbative predictions of the phase structure in a broad class of beyond-the-Standard-Model scenarios, including those featuring strong first-order phase transitions relevant for gravitational-wave signals, where conventional high-temperature approximations break down.
\end{abstract}

\maketitle

\section{Introduction}

Cosmological first-order phase transitions can provide the out-of-equilibrium conditions needed to generate the observed baryon asymmetry \cite{Sakharov:1967dj}, playing a central role in explaining the matter–antimatter imbalance of the Universe through electroweak (EW) baryogenesis \cite{Kuzmin:1985mm, Morrissey:2012db,Bodeker:2020ghk}. Such processes can also leave signatures in the form of a primordial gravitational wave (GW) background \cite{Hogan:1986dsh}, offering insight into the early Universe and revealing new physics beyond the Standard Model (BSM) \cite{Ramsey-Musolf:2019lsf}. Efforts to detect this GW background are ongoing \cite{Caprini:2015zlo,Caprini:2019egz,LISACosmologyWorkingGroup:2022jok}, with upcoming next-generation space-based experiments such as the Laser Interferometer Space Antenna (LISA) \cite{amaroseoane2017laserinterferometerspaceantenna} driving the search.

Predicting the GW signal from a cosmological phase transition driven by a scalar field in a BSM scenario is challenging due to the intricate thermal dynamics involved \cite{Arnold:1992rz,Farakos:1994kx,Braaten:1995cm}. The thermodynamics of such transitions are typically studied within the Euclidean imaginary-time formalism, where equilibrium properties at finite temperature $T$ are derived from a path integral defined in a compact time direction $\tau$ of size $\beta\equiv 1/T$, with fields featuring discretized frequency (Matsubara) modes (for a review, see Ref.~\cite{Laine:2016hma}). In principle, four-dimensional (4d) finite-$T$ lattice Monte Carlo simulations offer a fully non-perturbative approach to studying the phase structure and thermodynamic properties of such systems. In practice, however, long-standing challenges -- including the implementation of chiral fermions (see, e.g., Ref.~\cite{Luscher:2000hn}), difficulties in matching lattice and continuum formulations, and the significant computational cost of full 4d simulations with multiple fields -- render such studies largely impractical. Alternatively, turning to perturbative methods provides an approach devoid of such shortcomings, enabling controlled systematic improvements once higher loop corrections are available.

In perturbation theory, the equilibrium properties of phase transitions driven by an elementary scalar field $\phi$ are captured by the effective potential $V_{\text{eff}}(\varphi, T)$ -- an auxiliary quantity that encodes the phase structure of a quantum field theory (QFT) as a function of $T$ and a constant classical background field $\varphi$ for the scalar $\phi$. Local minima of the effective potential occurring at $\varphi_\text{min}=\langle\phi\rangle$ characterize the different phases the system may display, playing the role of a heuristic order parameter for the transition in question. Together with quantum corrections emerging in the zero-$T$ vacuum, the effective potential encodes how thermal effects can qualitatively modify the behavior of the classical tree-level potential $V_\text{tree}$, signaling the possibility of symmetry restoring phase transitions at high-$T$ \cite{PhysRevD.9.3320,PhysRevD.9.3357}.

More concretely, the effective potential can be determined from the expression (see Section 9.1 of Ref.~\cite{Laine:2016hma}):
\begin{equation}  
V_{\text{eff}}(\varphi, T) = V_{\text{tree}}(\varphi) - \frac{T}{\mathcal{V}} \ln \int \mathcal{D}[\Phi]\,e^{-\int_0^{\beta} \mathrm{d}\tau \int_\mathcal{V} L_\text{E}(\Phi)}, \label{effectivepotential}
\end{equation}
where $\mathcal{V}$ is the spatial volume of the system (here always taken to infinity), and the functional integration runs over a collection of fields $[\Phi]$ in any given QFT. Owing to the presence of a non-trivial background, the $\phi$ field now represents finite-momentum fluctuations around the new vacuum. This in turn alters the mass eigenvalues of the fields and couplings of the theory, information that is encoded in the parameters of the Euclidean Lagrangian $L_\text{E}$ entering the path-integral weight in Eq.~\eqref{effectivepotential}.

In the thermodynamic limit $\mathcal{V}\to\infty$, the effective potential -- when evaluated at the minimum corresponding to a given phase -- yields the equilibrium free-energy density $f(T)$.
Phase transitions occur at critical temperatures $T = T_c$, where the effective potential exhibits degenerate minima and $f(T)$ becomes non-analytic in $T$. If the first derivative, $f'(T)$, is discontinuous at $T=T_c$, the transition is classified as first-order.

From a field-theoretic perspective, the mass $M^2$ of light bosonic zero-frequency modes propagating in the thermal medium can receive perturbative corrections of the same order as $M^2$, thereby invalidating a naive expansion in small couplings within this regime. Approaching the transition point at $T_c$, these effects can become more severe due to masses becoming of non-perturbative order \cite{Laine:2000kv}. Fortunately, theoretical control can be restored upon implementing a suitable resummation scheme of light modes, allowing to systematically account for these effects. Physically, this breakdown of naive perturbation theory stems from the emergence of a hierarchical multi-scale system, making resummation an essential procedure for correctly capturing the non-trivial infrared (IR) dynamics of the theory \cite{Andersen:2004fp}.

The most successful approach to implement a proper resummation at high temperatures is an effective field theory (EFT) \cite{Kajantie:1995dw,Braaten:1995cm} based on dimensional reduction (DR) \cite{Ginsparg:1980ef,Appelquist:1981vg}. This framework has been extended to high loop orders in Quantum Chromodynamics (QCD) \cite{Kajantie:2002wa,Kajantie:2000iz,Kurkela:2016was} and the EW sector of the Standard Model (SM) \cite{Arnold:1992rz,Kajantie:1995kf,Kajantie:1996mn,Kajantie:1995dw,Farakos:1994kx,Farakos:1994xh}, while also proving valuable in various studies of BSM scenarios \cite{Bodeker:1996pc,Losada:1996ju,Cline:1996cr,Laine:1996ms,Andersen:1998br,Laine:1998wi,Laine:2000rm,Gynther:2005av,Gynther:2005dj,Brauner:2016fla,Andersen:2017ika,Gorda:2018hvi,Niemi:2018asa,Schicho:2021gca,Schicho:2022wty,Ekstedt:2022bff,Tenkanen:2022tly,Biondini:2022ggt}. DR relies on the existence of a high-$T$ scale hierarchy of the form $\pi T \gg gT \gg g^2T/\pi$ for a small coupling $g$ 
(for concreteness, we consider gauge theory), allowing to systematically integrate out modes characterized by the heavier scales. Since nonzero Matsubara modes and fermion fields carry heavy effective thermal masses of order $\pi T$, this procedure leaves a three-dimensional (3d) bosonic EFT where the sole propagating degrees of freedom are the light static Matsubara zero modes. In this theory, the presence of a soft $(gT)$ scale traces back to the perturbative Debye screening of electric fields, while the ultrasoft $(g^2T/\pi)$ scale is attributed to non-perturbative magnetic screening arising in non-Abelian interactions. While the 3d EFT describes the dynamics of light excitations only, the effects of the heavier scales are encoded in the coefficients of the effective operators, which can be determined perturbatively by demanding that long-distance static correlators match those of the full 4d theory \cite{Braaten:1995cm}. This ensures that all scales are properly accounted for, allowing the 3d EFT to correctly reproduce the resummed effective potential of the original 4d theory at high temperatures.

Focusing first on the minimal SM, where the EW sector is weakly coupled at the EW scale, 
DR provides a controlled framework for studying the phase transition accurately. In this case, the scale hierarchy at high temperature is well satisfied near the transition point, allowing both hard and soft modes to be perturbatively integrated out. The resulting 3d EFT is a super-renormalizable gauge-Higgs theory for ultrasoft modes in the high-$T$ confinement phase, which can be studied non-perturbatively on the lattice \cite{Farakos:1994xh,Laine:1995np,Laine:1997dy}. These analyses have reliably demonstrated that the SM transition is a smooth crossover rather than a first-order phase transition \cite{Kajantie:1995kf,Kajantie:1996mn}, hence not leaving behind gravitational waves.

However, in many BSM models, new degrees of freedom (including, e.g., scalar fields or fermions) couple to the Higgs field with strengths that are large enough to influence the thermal dynamics. This can lead to significant modifications of the low-energy effective theory, potentially resulting in stronger first-order phase transitions \cite{Niemi:2024axp}, and thus in GW signals that may fall into the sensitivity range of future detectors such as LISA \cite{Gould:2024jjt}. In attempts to precisely determine the underlying thermal parameters for the GW within these models -- such as phase transition strength, inverse duration, and transition completion temperature -- perturbative methods \cite{Croon:2020cgk,Gould:2021oba,Gould:2021ccf,Friedrich:2022cak,Tenkanen:2022tly,Kierkla:2023von,Lewicki:2024xan,Ramsey-Musolf:2024ykk,Chala:2024xll,Chala:2025aiz,Kierkla:2025qyz} and lattice simulations \cite{Moore:2000jw,Moore:2001vf,Gould:2019qek,Gould:2022ran,Gould:2024chm,Hallfors:2025key} within the 3d EFT framework are typically employed.
Extensive comparisons of perturbation theory and non-perturbative studies within the 3d EFT in gauge-Higgs systems \cite{Gould:2023ovu,Ekstedt:2022zro,Ekstedt:2024etx} have demonstrated that the mass scale associated with the transitioning scalar remains above the ultrasoft scale for strongly first-order phase transitions. This feature suppresses the emergence of the so-called Linde problem \cite{LINDE1980289,Braaten:1994na}, thereby ensuring the reliability of perturbation theory within the 3d EFT in describing the dynamics of the transition \cite{Kainulainen:2019kyp,Gould:2021dzl,Gould:2023ovu,Niemi:2024axp,Ekstedt:2024etx}.

Nonetheless, a pressing issue in all these studies is the validity of the 3d EFT description in BSM scenarios that exhibit strong phase transitions. Specifically, hypothetical degrees of freedom with large portal couplings $\lambda_\text{p}$ to the transitioning scalar field $\phi$ -- as required to trigger a first-order phase transition, in contrast to a crossover in the SM -- give rise to large masses $M \sim \sqrt{\lambda_{\text{p}}}\langle\phi\rangle \sim \pi T$ in the broken phase. This leads to large $O(1)$ effects from higher-order effective operators (see e.g., Refs.~\cite{Chala:2024xll,Bernardo:2025vkz,Chala:2025aiz,Chala:2025oul}) describing corrections to a high-$T$ asymptotic series in $M^2/(\pi T)^2$, an expansion which is often truncated to super-renormalizable order. As a result, the predictive power of such 3d EFT studies for strong transitions is put into question, highlighting the necessity of a framework that can not only accommodate large masses, but also further enables loop computations to be systematically extended to high orders. This would also allow to mitigate a sizable renormalization-scale dependence observed already at the state-of-the-art two-loop level in strong phase transitions~\cite{Laine:2017hdk}, a feature expected in the presence of large portal couplings.

In this work, we bridge this critical gap by developing a framework that not only encompasses the regime where high-$T$ expansions and non-perturbative 3d EFT methods remain valid, but also extends reliable predictions beyond it -- a crucial step for accurately describing the strong transitions most relevant for potential GW observations. Concretely, we introduce a novel perturbative approach to evaluate the full 4d resummed thermal (finite-$T$) effective potential across the entire mass range. By systematically disentangling the physics associated with hard- and soft-momentum scales, our framework enables their unified treatment within a single expression, thereby facilitating a clean separation and handling of divergences of both vacuum and thermal origin. At the practical level, the key advancement of our formulation lies in sorting contributions in a manner that allows an efficient numerical evaluation of massive multi-loop thermal sum-integrals \textit{without} any high-$T$ approximations. To this end, we apply for the first time a method capable of tackling such a demanding task, enabling the determination of the thermal effective potential with unprecedented accuracy. Furthermore, we anticipate that future works will greatly profit from these advances, proving pivotal in studies of thermal bubble nucleation \cite{Gould:2021ccf} aiming at going beyond high-$T$ expansions.

\section{A new framework}
\label{sec:framework}

The perturbative evaluation of the 4d thermal effective potential defined in Eq.~\eqref{effectivepotential} -- denoted from now on as $V^{\text{res}}_{\text{eff}}$ to emphasize the necessity of resummation -- without relying on high-$T$ expansions is a conceptually and technically non-trivial task. Two major challenges underlie this difficulty.

First, thermal resummations required to address soft physics become increasingly intricate at higher loop orders. Previous attempts rely on simplified schemes tailored for specific studies, often being impractical at higher orders or introducing ambiguities. Notable examples include the ``daisy resummation'' by Arnold and Espinosa \cite{Arnold:1992rz}, 
Parwani-resummation \cite{Parwani:1991gq}, various forms of screened perturbation theory, where the Lagrangian is augmented with effective thermal masses \cite{Karsch:1997gj}, among many others \cite{Boyd:1993tz,Chiku:1998kd,Chiku:2000eu,Kneur:2015uha,Funakubo:2012qc,Curtin:2016urg,Laine:2017hdk,Curtin:2022ovx,Funakubo:2023cyv,Funakubo:2023eic,Bahl:2024ykv,Bittar:2025lcr,Banerjee:1991fu,Boyanovsky:1996dc,Amelino-Camelia:1997xip,Pinto:1999py,Wainwright:2011qy,Kainulainen:2021eki,Funakubo:2023eic}. 

Second, standard (semi)analytic methods in thermal field theory struggle to evaluate dimensionally regulated multiloop sum-integrals involving massive propagators. So far, the simplest two‑loop cases have been worked out (see, e.g., Refs.~\cite{Laine:2000kv,Laine:2017hdk}), and their practical application to BSM models is hindered by medium‑dependent IR divergences that must be manually subtracted from the (semi)analytic results. Furthermore, whether these techniques are applicable beyond two loops is far from clear given the growing complexity of integral topologies. These points are essential for establishing a systematic determination of $V^{\text{res}}_{\text{eff}}$ beyond the high-$T$ expansion, and for assessing the issue of convergence once higher-order corrections are available.

With the aim of evaluating $V^{\text{res}}_{\text{eff}}$ without high-$T$ approximations, the aforementioned challenges can be addressed in a setting that conveniently reorganizes the perturbative series. Inspired by previous pressure computations in thermal QCD \cite{Kurkela:2016was}, this formulation incorporates a simple yet powerful separation of hard and soft mode contributions, expressed as
\begin{equation}
\begin{split}
V^{\text{res}}_{\text{eff}} & = V^{\text{res}}_{\text{eff}} - V^{\text{res,soft}}_{\text{eff}} + V^{\text{res,soft}}_{\text{eff}}\\
& \simeq   \left (V^{\text{naive}}_{\text{eff}} - V^{\text{naive,soft}}_{\text{eff}}\right )  + V^{\text{res,soft}}_{\text{eff}}.
\end{split}
\label{eq:mastereq}
\end{equation}
In the first step, we add and subtract a function $V^{\text{res,soft}}_{\text{eff}}$, whose purpose is to implement a proper resummation of all soft modes in need of nonperturbative treatment. This ensures that the difference $V^{\text{res}}_{\text{eff}} - V^{\text{res,soft}}_{\text{eff}}$ captures contributions from heavy hard modes only, implying that this quantity is insensitive to IR effects and can therefore be determined within a naive (i.e., unresummed) perturbative loop expansion.  Note that if any hard contributions are inadvertently included in the precise definition of $V^{\text{res,soft}}_{\text{eff}}$, they are automatically subtracted by $V^{\text{naive,soft}}_{\text{eff}}$, thus preventing potential overcounting. 

Examining the effective potential in Eq.~\eqref{effectivepotential} further, the zero-$T$ limit encodes how quantum corrections modify the classical tree-level potential, which amounts to the well-known Coleman-Weinberg (CW) mechanism inducing radiative symmetry breaking in vacuum QFT \cite{Coleman:1973jx}. This contribution is captured by a naive weak-coupling expansion in $V_\text{eff}^\text{naive}$, while only thermal pieces require resummation. As a result, vacuum and thermal parts play distinct roles in our discussion of thermal phase transitions, leading us to split the effective potential in two pieces
\begin{equation}
    V_\text{eff}^\text{res}(\varphi,T)=\underbrace{(V_\text{tree}+V^\text{CW})}_{T=0}+\underbrace{V_{T,\text{eff}}^\text{res}}_{T\neq0}, \label{eq:TzeroTnonzero}
\end{equation}
where $V_{T,\text{eff}}^\text{res}$ is the temperature-dependent contribution sensitive to IR effects, for which the reorganization displayed in Eq.~\eqref{eq:mastereq} is conducted (note that soft contributions are purely thermal). In contrast to $V_{T,\text{eff}}^\text{res}$, the vacuum part is amenable to highly developed methods in QFT, enabling the computation of $V^\text{CW}$ -- consisting of massive vacuum bubbles -- analytically to a high loop order (see e.g., Ref.~\cite{Luthe:2016spi}).

Upon rearranging contributions in Eq.~\eqref{eq:TzeroTnonzero} and implementing thermal resummations in Eq.~\eqref{eq:mastereq}, divergences of distinct origin are featured in the expressions, warranting further inspection of how the cancellation plays out in the full resummed effective potential. In the case of zero-$T$ contributions, background-dependent ultraviolet (UV) divergences in $V^\text{CW}$ are removed by replacing couplings and masses in $V_\text{tree}$ and lower-order terms for the corresponding renormalized quantities. A residual divergent (background-independent) vacuum contribution remains in the effective potential after renormalization; however, as usual, it cancels out in physical observables. For the thermal part, on the other hand, cancellations occur among the three terms in Eq.~\eqref{eq:mastereq} in the following manner: 
\begin{itemize}
\item First, UV divergences in $V_{T,\text{eff}}^\text{res}$ are removed by renormalization of 4d parameters in lower-order contributions.

\item Second, IR divergences in $V_{T,\text{eff}}^\text{naive}$ are eliminated by those in $V^\text{naive,soft}_\text{$T$,eff}$, leaving residual UV divergences.

\item And third, leftover UV divergences are seen to cancel out against $V^\text{res,soft}_\text{$T$,eff}$, a quantity that is IR safe by construction. 

\end{itemize}
In what follows, we keep vacuum and thermal parts together in $V^\text{res}_\text{eff}$, bearing in mind the interplay of UV and IR divergences in the different contributions.

Concentrating on the soft terms in Eq.~\eqref{eq:mastereq}, a minimal yet effective description of $V^{\text{res,soft}}_{\text{eff}}$ is to resum the static light bosonic modes within the framework of dimensional reduction. Importantly, to accurately describe the effective potential across the whole range of background-field values, the construction of the EFT by demanding correlators match between the EFT and the full 4d theory \emph{must be carried out without relying on high-$T$ expansions}. This, in turn, ensures that $V^{\text{naive,soft}}_{\text{eff}}$ 
systematically cancels all IR-sensitive contributions in $V^{\text{naive}}_{\text{eff}}$ at all orders, and crucially, does so locally at the integrand level.

With the aim of numerically evaluating thermal sum-integrals in the following, we define the IR safe quantity $\Delta V^\text{hard}_\text{eff}\equiv V^{\text{naive}}_{\text{eff}} - V^{\text{naive,soft}}_{\text{eff}}$. We note in passing that this quantity closely resembles the definition of the matching coefficient associated with the unit operator in the EFT, as originally introduced by Braaten and Nieto in the context of the free energy density in QCD \cite{Braaten:1995cm,Braaten:1995jr}.

To evaluate the $\Delta V^{\text{hard}}_{\text{eff}}$ contribution in \cref{eq:mastereq}, which becomes increasingly challenging at higher loop orders, we employ a novel algorithmic technique: a finite-temperature extension of Loop-Tree Duality (LTD), referred to as \texttt{hotLTD}. This method enables the direct numerical evaluation of multiloop sum-integrals with massive propagators in momentum space \cite{Capatti}. Originally developed to render vacuum Feynman integrals numerically tractable via Monte Carlo methods (see, e.g., Refs.~\cite{Soper:1999xk,Catani:2008xa,Capatti:2019ypt,Capatti:2019edf,Capatti:2020ytd,Capatti:2020xjc,Capatti:2022tit,Buchta:2015wna,Becker:2012bi,Runkel:2019yrs}), LTD has recently been extended to a finite-density setting \cite{Navarrete:2024zgz,Karkkainen:2025nkz}. Building on these developments, \texttt{hotLTD} enables the high-precision computation of $\Delta V^{\text{hard}}_{\text{eff}}$, even beyond two loops and without relying on the high-$T$ expansion. The procedure involves: (i) algorithmic subtraction of UV divergences through an integrand-level implementation of the Bogoliubov's $R$-operation \cite{Bogoliubov:1957gp,Hepp:1966eg,Zimmermann:1969jj} (see also Refs.~\cite{Capatti:2019edf,Capatti:2022tit} for a more modern exposition); (ii) analytic evaluation of Matsubara sums, yielding a closed-form three-dimensional integrand; and (iii) Monte Carlo integration of the resulting spatial integral. A key remaining challenge is the systematic subtraction of temperature-dependent IR divergences, which we address in \cref{sec:threeloop}. A complete two-loop example is provided in \cref{app:hotLTD-example}.

\section{A proof-of-principle demonstration}
\label{sec:pop}

To establish a proof-of-principle demonstration of \eq\eqref{eq:mastereq} in a concrete field theory setting, we consider a scalar-Yukawa theory consisting of a real scalar field $\phi$ and a Dirac fermion $\psi$. This simple model serves as an effective proxy for any scalar-driven first-order cosmological phase transition, including the EW phase transition, and allows us to introduce our novel framework in a minimal setting. Whereas in gauge-Higgs systems a phase transition is triggered by a scale hierarchy between the light scalar undergoing the transition and a heavier gauge field or additional scalar field \cite{Hirvonen:2022jba,Gould:2023ovu}, in the scalar-Yukawa model with a transitioning field $\phi$, it is the fermion $\psi$ that acts as the heavy field inducing the transition.

The bare Euclidean Lagrangian for the model is \cite{Andersen:1997hq}
\begin{equation}
\begin{split}    
    L_{\text{E}} &=\frac{1}{2}(\partial_\mu\phi)^2+\sigma\phi+\frac{1}{2}m^2_\phi \phi^2+\frac{1}{3!}g\phi^3+\frac{1}{4!}\lambda\phi^4  \\
    &+ \bar{\psi}(\slashed{\partial}+m_\psi) \psi+y\phi\bar{\psi}\psi, 
\end{split}   
    \label{yuk2}
\end{equation}
where the cubic coupling has mass dimension one, and the tadpole parameter has mass dimension three. To construct the effective potential as in Eq.~\eqref{effectivepotential}, we may simply shift the scalar field by a classical background, $\phi \to \phi + \varphi$. This effectively singles out quantum fluctuations contributing to the path integral, resulting in the corresponding fermion- and scalar-mass eigenvalues
\begin{align}
\label{eq:masses}
M^2_\phi(\varphi) &= m^2_\phi + g \varphi + \frac12 \lambda \varphi^2, \\
\label{eq:masses2}
M_\psi(\varphi) &= m_\psi + y \varphi,
\end{align}  
while the cubic vertex coupling becomes 
\begin{equation}
G(\varphi) = g + \lambda \varphi.
\label{eq:cubicG}
\end{equation}
Meanwhile, linear terms in the fluctuation integrate to zero in the path-integral weight in Eq.~\eqref{effectivepotential}, delegating the role played by the tadpole term to just the tree-level potential. 

Following Ref.~\cite{Gould:2023jbz}, we adopt a power counting scheme in which the loop expansion parameters in vacuum are of similar magnitude, i.e., $g^2/m^2 \sim y^2 \sim \lambda$ for both vacuum masses $m_\phi$ and $m_\psi$. Additionally, a phase transition triggered at high temperature reflects the effect of thermal corrections qualitatively modifying the potential in vacuum. In view of these considerations, we demand the thermal corrections to be comparable to contributions from the tree-level potential, leading us to further establish $\sigma/T^2 \sim g$ and $m^2/T^2 \sim \lambda$ within our scheme.

Turning to the equilibrium thermodynamics of this model, extensive studies have been carried out in Refs.~\cite{Gould:2021dzl,Gould:2023jbz} (see also Refs.~\cite{Chala:2019rfk,Chala:2025aiz}), where the masses $M^2_\phi(\varphi)$ and $M^2_\psi(\varphi)$ appearing in propagators are always expanded at high temperatures. Under this approximation, these studies have explored both non-perturbative lattice Monte Carlo simulations and renormalization-group improved perturbation theory up to three-loop level \cite{Gould:2021dzl,Rajantie:1996np}, where 
both approaches involve computing massless thermal sum-integrals in the dimensional reduction stage. In what follows, we relax the assumption of a high-$T$ expansion and aim to compute the resummed thermal 4d effective potential by means of Eq.~\eqref{eq:mastereq}, up to two loops, in terms of the full background-field-dependent masses displayed in Eq.~\eqref{eq:masses}. Furthermore, we illustrate -- through a particularly instructive three-loop example -- how this computation can be extended beyond the two-loop level.

Concretely, for our toy model we may write
\begin{equation} 
V^\text{res}_\text{eff}(\varphi,T) = V_\text{tree} + V^{(1)\text{res}}_\text{eff} + V^{(2)\text{res}}_\text{eff} + O(\text{3-loop}), \label{Vreseff}
\end{equation}
where the tree-level potential reads
\begin{equation}
    V_\text{tree}=\sigma \varphi+\frac{1}{2}m_\phi^2\varphi^2+\frac{1}{3!}g\varphi^3+\frac{1}{4!}\lambda\varphi^4, \label{eq:treelevel}
\end{equation}
and the superscripts in Eq.~\eqref{Vreseff} denote the corresponding loop correction. To evaluate this expansion at each order, we perform the following three-step procedure.

First, we generate the (bare) naive contribution $V^{\text{naive}}_{\text{eff}}$ by performing the usual loop expansion in the full theory, a task most conveniently automated to high loop orders with \texttt{Qgraf} \cite{Nogueira:1991ex} and the computer algebra package \texttt{FORM} \cite{Vermaseren:2000nd}. To two loops, this exercise yields
\begin{widetext}
\begin{align}
    -V^{(1)\text{naive}}_\text{eff} &= \raisebox{-0.4\height}{\includegraphics[scale=0.8]{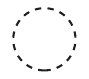}}+\raisebox{-0.4\height}{\includegraphics[scale=0.8]{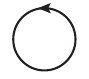}}=\frac{1}{2}\SumInt_{P}\ln\left(P^2+M_\phi^2\right)-4\times \frac{1}{2}\SumInt_{\widetilde{P}}\ln\left(P^2+M_\psi^2\right),  \label{eq:naive1}\\
    -V^{(2)\text{naive}}_\text{eff} &= \frac{1}{8}\raisebox{-0.4\height}{\includegraphics[scale=0.8]{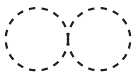}}-\frac{1}{2}\raisebox{-0.4\height}{\includegraphics[scale=0.8]{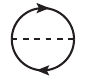}}+\frac{1}{12}\raisebox{-0.4\height}{\includegraphics[scale=0.8]{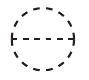}} \nonumber \\
    &= -\frac{\lambda}{8} [\mathcal{I}(M_\phi)]^2 +y^2\Big\{(M^2_\phi-4M^2_\psi)\widetilde{\mathcal{S}}(M_\psi,M_\phi)+2\mathcal{I}(M_\phi)\widetilde{\mathcal{I}}(M_\psi)-[\widetilde{\mathcal{I}}(M_\psi)]^2\Big\} +\frac{G^2}{12}\mathcal{S}(M_\phi),
    \label{eq:naive2}
\end{align}
\end{widetext}
where dashed (solid) lines represent scalar (fermion) fields. In Appendix \ref{app:details}, the corresponding definitions for the sum-integrals can be found; the calligraphic $\mathcal{I}$ and $\mathcal{S}$ denote one- and two-loop massive bosonic sum-integrals, respectively, with a wavy line on top indicating fermionic signature of the corresponding integration measure.

Second, the $V^{\text{res,soft}}_{\text{eff}}$ contribution is generated from the path integral defined in the 3d effective theory,
\begin{equation}
\begin{split}
    V_\text{eff}^\text{res,soft}&(\varphi_3,T)=-\!\!\lim_{\mathcal{V}\to \infty} \frac{T}{\mathcal{V}} \ln \int \!\mathcal{D}\phi_3\, e^{-\!\int_\mathcal{V} \, L^\text{3d}_\text{E}(\phi_3)}.
    \label{eq:softres}
\end{split}
\end{equation}
Here, the EFT describes 3d light bosonic fluctuations $\phi_3$ over a background $\varphi_3=\varphi/\sqrt{T}$, and is defined by the 3d Euclidean Lagrangian
\begin{align}
L^\text{3d}_\text{E}=\frac{1}{2}(\partial_i\phi_3)^2+\frac{1}{2}M^2_{3}\phi_3^2+\frac{G_3}{3!}\phi_3^3+\frac{\lambda_3}{4!}\phi_3^4 +\delta L, \label{3dlagrangian}
\end{align}
where $\delta L$ represents an infinite tower of higher-dimensional operators that encode higher-order effects in our perturbative expansion. In a setting where all mass effects are retained through fully massive sum-integrals -- i.e., without resorting to high-$T$ expansions -- the impact of higher-dimensional operators naturally aligns with the coupling expansion. Therefore, at the perturbative order we consider, it suffices to truncate the EFT at super-renormalizable order, corresponding to $\delta L = 0$. At this level, the effective parameters entering Eq.~\eqref{3dlagrangian} -- specifically $M_3^2$ and $G_3$ -- depend on the 3d background field and parameters in identical form compared to the full theory (see Eqs.~\eqref{eq:masses} and \eqref{eq:cubicG}), and receive loop corrections from the heavy scales through a naive perturbative matching of static $n$-point Green's functions between the effective and full theories \cite{Kajantie:1995dw}. This allows to solve for the effective parameters in terms of $T$ and the full-theory parameters, up to the desired order in the loop-expansion couplings $\lambda$, $g$ and $y^2$. 

Collectively denoting the effective parameters by $X_3$, their generic form is $X_3=X+\delta X_3$, where $X$ are the corresponding 4d parameters scaled by trivial $T$ factors (to account for correct mass dimensions) and $\delta X_3$ are loop corrections. These loop contributions are IR finite by construction, encoding the effects of the hard modes that were integrated out. Within our model, they have been computed to high orders only in a high-$T$ expansion \cite{Gould:2023jbz}, requiring a generalization to the full non-expanded case. The required cases for our paper are collected in Appendix \ref{app:matchingparameters}.

A diagrammatic expansion of \eq\nr{eq:softres} results in integrals featuring resummed propagators:
\begin{align}
V^{(1)\text{res,soft}}_\text{eff} & = \frac{T}{2}\int_\mathbf{p} \ln\left(p^2+M_{3}^2\right), \label{Veffressoft1} \\
V^{(2)\text{res,soft}}_\text{eff} & = -\frac{(G_3)^2}{12T}S_{3d}(M_{3}) + \frac{\lambda_3}{8T}[I_{3d}(M_{3})]^2, \label{Veffressoft2}
\end{align}
where the results for the corresponding massive one- and two-loop 3d integrals are collected in Appendix~\ref{app:details}. Here, the mass parameter $M_3^2$ is required to one-loop accuracy (see Appendix~\ref{app:effectivemass} for details), whereas the effective couplings are needed only at leading order (LO), namely $G_3= G/\sqrt{T}$ and $\lambda_3= \lambda T$.

The third and final step in our evaluation of Eq.~\eqref{Vreseff} refers to the construction of the $V^{\text{naive,soft}}_{\text{eff}}$ contribution. This object aims to capture the IR behavior of the theory in an unresummed fashion, and can be obtained directly from the diagrammatic evaluation of $V^{\text{res,soft}}_{\text{eff}}$ -- the result of which is displayed in Eqs.~\eqref{Veffressoft1}-\eqref{Veffressoft2} -- upon an expansion of all effective parameters $X_3$ in powers of $\delta X$, up to the order dictated by our coupling expansion. This yields:
\begin{align}
    V^{(1)\text{naive,soft}}_\text{eff} &= \frac{T}{2}\int_\mathbf{p} \ln\left(p^2+M_\phi^2\right), \label{Veffnaivesoft1} \\
    V^{(2)\text{naive,soft}}_\text{eff} &= \frac{\lambda}{8}[I_\text{3d}(M_\phi)]^2 - \frac{G^2}{12}S_\text{3d}(M^2_\phi) \nonumber \\
    &+ \frac{1}{2}(\delta M^2_3)I_\text{3d}(M_\phi). \label{Veffnaivesoft2}
\end{align}
Here, the second line of Eq.~\eqref{Veffnaivesoft2} originates from expanding the mass parameter in the one-loop resummed expression in Eq.~\eqref{Veffressoft1} to next-to-leading order (NLO). The remaining contributions correspond to one- and two-loop diagrams in the EFT keeping effective parameters at leading order -- in particular, the mass parameter appearing in the now unresummed propagators.

To understand the role played by the different terms generated in these steps -- and more generally in Eq.~\eqref{eq:mastereq} -- we now turn to inspect the simple case of the full resummed effective potential to one-loop order. It reads
\begin{align}
    V^{\text{res}}_\text{eff}(\varphi,T) &\simeq V_\text{tree}+ \Delta V^{(1)\text{hard}}_\text{eff} + V^{(1)\text{res,soft}}_\text{eff}, \label{masterequation1loop}
\end{align}
where the tree-level and one-loop resummed potentials are shown in Eqs.~\eqref{eq:treelevel} and \eqref{Veffressoft1}, respectively, while the one-loop hard contribution $\Delta V^{(1)\text{hard}}_\text{eff}\equiv V^{(1)\text{naive}}_{\text{eff}} - V^{(1)\text{naive,soft}}_{\text{eff}}$ can be written in the form
\begin{widetext}    
\begin{equation}   
\begin{split}
       \Delta V^{(1)\text{hard}}_\text{eff}  & = 
    \left ( \frac{1}{2}\SumInt_{P}\ln\left( P^2+M^2_\phi \right) - \frac{T}{2}\int_\mathbf{p} \ln\left(p^2+M_\phi^2\right)\right ) - 4\times\frac{1}{2}\SumInt_{\widetilde{P}}\ln\left( P^2+M^2_\psi\right). \label{1loopveff}
\end{split}
\end{equation}
\end{widetext}
Here, the 3d integral present in $\Delta V^{(1)\text{hard}}_\text{eff}$ corresponds to that obtained in Eq.~\eqref{Veffnaivesoft1}. This integral has the effect of removing the massless zero mode from the full bosonic sum-integral, which along with its fermionic counterpart -- an object free of massless modes in the sum -- results in only hard modes with effective masses $\sim\pi T$ contributing to $\Delta V^{(1)\text{hard}}_\text{eff}$. On the other hand, $V^{(1)\text{res,soft}}_\text{eff}$ in Eq.~\eqref{Veffressoft1} properly implements the resummation of the zero-mode contribution at one-loop order.

While the (divergent) vacuum ($T=0$) contribution of the two massive sum-integrals entering Eq.~\eqref{1loopveff} can be easily solved analytically (see Eq.~\eqref{appeq:logvacuum} for the result), the thermal parts are non-trivial (finite) functions of $M_\phi/T$ and $M_\psi/T$. These functions can, however, be expressed in terms of the simple one-dimensional numerical integral $h$ defined in Eq.~\eqref{appeq:hfunction}. The remaining 3d integral in Eq.~\eqref{1loopveff} is, in contrast, trivial to solve analytically, and the result can be found in Eq.~\eqref{3dintegrals1loop}.

As outlined in the previous section, the removal of vacuum UV divergences in the one-loop correction is achieved upon replacing the parameters in the tree-level potential for the renormalized ones (the relevant counterterms are collected in Appendix~\ref{sec:betas}). Working in the $\overline{\text{MS}}$ scheme, we write the renormalized full one-loop resummed effective potential as
\begin{align}
    V^\text{res}_\text{eff,R}(\varphi,T)=V_\text{tree,R}+V^\text{(1)res}_\text{eff,R}+O(\text{2-loop}), \label{eq:veffren}
\end{align}
where the subscript R refers to quantities expressed in terms of the (running) renormalized parameters, and the one-loop correction reads
\begin{widetext}
\begin{equation}
\begin{split}
\label{eq:Veffren1loop}
    V^\text{(1)res}_\text{eff,R}(\varphi,T) &= \frac{M_\psi^4}{(4\pi)^2}\left( \ln\frac{\overline{\Lambda}^2}{M^2_\psi}+\frac{3}{2} \right) -\frac{M_\phi^4}{4(4\pi)^2}\left( \ln\frac{\overline{\Lambda}^2}{M^2_\phi}+\frac{3}{2} \right) + \frac{T^4}{2\pi^2}\left[ h(q_\phi)-\frac{1}{2}h(2q_\psi)+4h(q_\psi) \right] \\
    & -\frac{T}{12\pi} \left[ (M^2_3)^{3/2}  - (M^2_\phi)^{3/2} \right] 
    + O(\text{2-loop}).
\end{split}
\end{equation}
\end{widetext}
Here, $\Lbar$ stands for the (4d) renormalization scale, and $q\equiv M/T$ represents the dimensionless ratio of the bosonic or fermionic mass eigenvalue to the temperature. By making use of the one-loop beta functions from Appendix~\ref{sec:betas}, we have checked the cancellation of the explicit logarithms of $\Lbar$ with the renormalization-group running of the tree-level potential. Furthermore, note that in Eq.~\eqref{eq:Veffren1loop} we have expressed the fermionic thermal integral in terms of the bosonic $h$ function, a convenient relation obtained simply by rescaling the loop momentum and splitting the Matsubara sum in odd and even modes.

We now make a few observations regarding Eq.~\eqref{eq:Veffren1loop}. First, the last two terms amount to the zero-mode resummed and unresummed soft contributions, featuring the familiar non-analytic behavior in the masses. Second, for values in the parameter-space region where $M_\phi^2\sim \pi T$ in the broken phase -- regime where a high-$T$ expansion is not justified -- the soft terms in the second line cancel up to higher-order sub-leading loop corrections to $M^2_\phi$. This depicts a situation where the hard-scale physics dominates the thermal dynamics, enabling one to accurately describe the effective potential within a naive loop expansion. Third, in the case of soft scalar masses -- e.g., when $M_\phi^2\sim\lambda T^2$, a regime where a high-$T$ expansion is legitimate -- the loop corrections induced by hard modes are of the same order as $M_\phi^2$. This indicates that the resummed soft contribution plays a significant role, highlighting the necessity of resummation for an accurate description of the thermodynamics. In this regime, the terms on the second line of \eq\nr{eq:Veffren1loop} coincide with the commonly adopted ``daisy resummation'' by Arnold and Espinosa \cite{Arnold:1992rz}. However, we emphasize -- in contrast to the extensive literature on daisy resummations (c.f. e.g.~\cite{Delaunay:2007wb,Espinosa:2011ax}) -- that the thermal mass $M_3^2$ appearing in our resummed contribution is computed entirely without resorting to high-$T$ expansions. This constitutes a novel addition to our framework, which we demonstrate below to be critical for maintaining the consistency of resummations at higher orders.

At this stage, we find it pertinent to address a persistent conceptual -- and practical -- issue concerning the perturbative computation of the effective potential: the emergence of imaginary parts associated with tachyonic masses $M^2_\phi<0$. Although at zero temperature such contributions are typically interpreted as decay rates of unstable phases \cite{Weinberg:1987vp,Andreassen:2016cvx}, within our framework we can shed new light on this point. Concentrating on our one-loop result in Eq.~\eqref{eq:Veffren1loop} for concreteness, two sources of imaginary parts are seemingly present: logarithms and fractional powers of squared masses. The former terms, originating from zero-temperature divergences, cancel against the corresponding thermal contributions in the full expressions, leaving behind single logarithms of the form $\log(\overline{\Lambda}/2\pi T)$. Although such cancellation is trivially implemented within a high-$T$ expansion of the sum-integrals, in our full computation it takes place already at the integrand level -- via an exact cancellation between vacuum contributions and thermal distribution functions -- in the kinematic regime where bosonic energies are imaginary. The remaining real parts often exhibit singularities at spatial momenta $\mathbf{p}^2=m^2$, where the energy vanishes and the Bose-Einstein distribution diverges -- reflecting the typical infrared behavior of thermal sum-integrals. As a result, this singular structure is removed entirely by zero-mode contributions, which are responsible for generating the characteristic fractional powers of squared masses.

Taken together, all contributions carrying imaginary parts in the naive loop expansion are fully eliminated -- either internally within each sum-integral or through cancellations with the corresponding thermal counterterms in $V^\text{naive,soft}_\text{eff}$. Consequently, $\Delta V^\text{hard}_\text{eff}$ remains a perfectly real and finite quantity, not only in the massless limit but also throughout the tachyonic regime $M^2_\phi<0$ (these extensions are described in detail in Appendix~\ref{app:details}). The sole remaining potential source of imaginary components is the resummed zero-mode contribution $V^\text{res,soft}_\text{eff}$, such as the non-analytic $(M_3^2)^{3/2}$ term in Eq.~\eqref{eq:Veffren1loop}. However, even before entering the tachyonic regime, small positive values of $M_3^2$ -- possibly of non-perturbative magnitude -- may suggest the possibility of loop effects driving the thermal mass away from ever reaching this regime. Since nothing prevents a tower of higher-order corrections from changing this conclusion, we interpret the emergence of imaginary components as a signal of breakdown in the perturbative resummation scheme. Fortunately, in first-order phase transitions, such effects arise only near the potential barrier -- far from the physical minima -- leaving thermodynamic observables largely unaffected, a point we verify in all our computations. We stress that these important considerations are frequently neglected in the literature, with some approaches sidestepping the issue by either taking the real part of the potential or by simply replacing squared masses with their absolute values -- typically with no further justification.

\section{Two-loop effective potential and cancellation of IR divergences}

Proceeding to two loops, by collecting the diagrammatic expansions derived in the previous section for the various contributions to the (bare) effective potential -- namely, Eqs.~\eqref{eq:naive2}, \eqref{Veffressoft2} and \eqref{Veffnaivesoft2} -- the contributions can be arranged in a manner analogous to the one-loop case. Along with the resummed soft piece from Eq.~\eqref{Veffressoft2}, the two-loop hard contribution can be written in the form
\begin{widetext}
\begin{equation}
\begin{split}
\label{eq:veffhard2}
    \Delta V_\text{eff}^\text{(2)hard}&= -\frac{G^2}{12}\left [ \mathcal{S}(M_\phi)-S_\text{3d}(M_\phi) \right ] + \frac{\lambda}{8}\left\{ [\mathcal{I}(M_\phi)]^2-[I_\text{3d}(M_\phi)]^2 \right\}-\frac{1}{2}(\delta M^2_3)I_\text{3d}(M_\phi) \\
    &-y^2\left\{ (M^2_\phi-4 M^2_\psi)\widetilde{\mathcal{S}}(M_\psi,M_\phi)+2\mathcal{I}(M_\phi)\widetilde{\mathcal{I}}(M_\psi)-[\widetilde{\mathcal{I}}(M_\psi)]^2 \right\}.
\end{split}
\end{equation}
\end{widetext}
While this expression resembles the one-loop correction in Eq.~\eqref{1loopveff}, key differences arise at two-loop order. As in the one-loop result, the 3d integrals appearing in the first two terms of Eq.~\eqref{eq:veffhard2} explicitly subtract the zero-mode contributions from the corresponding fully bosonic sum-integrals. Crucially, and in contrast to the one-loop case, the presence of a certain sum-integral -- namely, the bosonic ``sunset'' $\mathcal{S}$ defined in Eq.~\eqref{app:bosonicsunset} -- renders the naive loop expansion IR divergent in the limit of vanishing boson masses. In our definition of $\Delta V_\text{eff}^\text{hard}$, however, this logarithmic IR sensitivity is explicitly removed by $S_\text{3d}$ at integrand level, highlighting the crucial role of the EFT construction in providing the appropriate subtractions.

Furthermore, beginning at two-loop order, contributions involving mixed zero- and non-zero-mode components may exhibit additional sensitivity to soft infrared effects. This sensitivity is accounted for by terms that emerge upon expanding effective parameters in resummed integrals, corresponding to the last term on the first line in Eq.~\eqref{eq:veffhard2}. Indeed, the effective mass correction $\delta M^2_3$ encapsulates the contribution of hard modes running in loops within the scalar two-point function evaluated at soft external momentum (see Appendix~\ref{app:effectivemass} for details). These contributions correspond precisely to the infrared limit of one-loop subgraphs -- i.e., expanded in small external momentum -- embedded within the full two-loop diagrams contributing to the effective potential, thereby eliminating any residual IR sensitivity. While not strictly required at two-loop order in the case of positive squared masses -- owing to the absence of diagrams exhibiting IR-sensitive mixed-mode contributions, an example of which will be presented in the next section -- this mechanism becomes essential at three loops and beyond. However, as noted at the end of the previous section, extending the definition of sum-integrals to the regime $M^2_\phi<0$ may introduce additional singular behavior akin to infrared divergences. Consequently, in this regime, mixed-mode contributions must be treated consistently at the integrand level in our numerical evaluation of thermal functions (see Appendix~\ref{app:details} for further details).

Turning to the explicit computation of the two-loop effective potential, one finds it to be significantly more involved than at one-loop order. Remarkably, however, the required non-trivial two-loop sum-integrals -- namely, the sunsets $\mathcal{S}$ and $\widetilde{\mathcal{S}}$ -- remain tractable using traditional methods \cite{Laine:2017hdk}. Nevertheless, in order to systematically extend these computations to higher orders for generic sum-integral structures, we adopt a numerical evaluation based on the \texttt{hotLTD} technique. A demonstration of the different steps involved in our computation is presented with more detail for the IR-sensitive $\mathcal{S}$ in Appendix~\ref{app:hotLTD-example}, 
while the final expressions for both sunset sum-integrals are presented in Eq.~\eqref{app:sunsetresults}.

Collecting our results for all (sum-)integrals in Eq.~\eqref{eq:veffhard2} from Appendix~\ref{app:details}, we perform the renormalization at two-loop level by means of the counterterms listed in Appendix~\ref{sec:betas}, witnessing a proper cancellation of all divergences. The resulting expression for the two-loop correction to the resummed effective potential in Eq.~\eqref{eq:veffren} then yields
\begin{widetext}
\begin{equation}
\begin{split}
\label{eq:veffren2loop}
    V^\text{(2)res}_\text{eff,R}(\varphi,T)&= \frac{\lambda M^4_\phi}{(4\pi)^4}B_1\left(\Lbar/M_\phi\right)+\frac{G^2 M^2_\phi}{(4\pi)^4}B_2\left(\Lbar/M_\phi\right)+\frac{y^2 M^4_\psi}{(4\pi)^4}F\left(\Lbar/M_\psi,\Lbar/M_\phi\right) \\
    &+\frac{\lambda T^4}{(4\pi)^4}\left\{ 8r^2-2rq_\phi^2 \left( \ln\frac{\Lbar^2}{M_\phi^2}+1 \right) -2\pi^2(q_\phi^2-q_3^2) \right\} + \frac{T^4}{8\pi}(q^2_\phi)^{1/2}(\delta q^2_\phi) \\
    &+\frac{16y^2T^4}{(4\pi)^4}\left\{ 4 \widetilde{r}^2-8r\widetilde{r}+ \widetilde{r}q^2_\phi \left( \ln\frac{\Lbar^2}{M^2_\phi}+1 \right) + q^2_\psi \left(r-\widetilde{r}\right)\left( \ln\frac{\Lbar^2}{M^2_\psi}+1 \right) \right. \\
    &\left. -(q^2_\phi-4q^2_\psi)\left[ \left(2\widetilde{r}+r\right)\ln\frac{\Lbar}{2\pi T} + 16\pi^4f_T \right] \right\} - \frac{4G^2 T^2}{(4\pi)^4}\left\{ r\ln\frac{\Lbar}{2\pi T}-\frac{\pi^2}{3}\left[\ln\frac{q_3}{2\pi}-16\pi^2b_T\right]\right\}.
\end{split}
\end{equation}
\end{widetext}
Here, the functions $B_1$, $B_2$, and $F$, as given in Eq.~\eqref{app:twoloopCW}, encapsulate the zero-$T$ part of the effective potential. In addition, we define $q_3\equiv M_3/T$ as the dimensionless 3d mass parameter, while the functions $r=r(q_\phi)$ and $\widetilde{r}=\widetilde{r}(q_\psi)$ denote the bosonic and fermionic thermal contributions to the basic one-loop tadpole sum-integral, respectively (see Eq.~\eqref{appeq:tadpole1}; and Eq.~\eqref{appeq:fulltadpoles} for the tachyonic extension). The two-loop thermal functions $b_T=b_T(q_\phi)$ and $f_T=f_T(q_\psi,q_\phi)$, introduced in Eq.~\eqref{app:sunsetresults}, are evaluated numerically via the \texttt{hotLTD} technique \cite{ZenodoData}. Furthermore, to check the physical consistency of our two-loop computation, we have witnessed the cancellation of all explicit logarithms of the renormalization scale in Eq.~\eqref{eq:veffren2loop} with the renormalization-group running of lower-order contributions, as dictated by the beta functions listed in Appendix~\ref{sec:betas}.

Notably, all contributions in Eq.~\eqref{eq:veffren2loop} interpolate smoothly to the regime where high-$T$ expansions are valid and remain finite in the exact massless limit. Moreover, the extension to the $M^2_\phi<0$ domain -- most critically for the two-loop thermal functions $b_T$ and $f_T$ -- is achieved by treating all terms generated in $V^\text{naive,soft}_\text{eff}$ on equal footing with the full sum-integrals at the integrand level. Extending the discussion from the previous section to the two-loop level, we highlight the role of the novel contribution $(\delta M^2_3)I_\text{3d}$ in cancelling non-analytic terms in $M^2_\phi$ that arise from mixed-mode contributions -- most notably the so-called "linear terms" (i.e., proportional to $M_\phi$), which historically motivated thermal resummation due to their bad perturbative behavior \cite{Shaposhnikov:1991cu,Dine:1992wr}. We emphasize that all terms in $V^\text{naive,soft}_\text{eff}$ must be regarded as genuine thermal counterterms and always consistently incorporated at the integrand level alongside the full sum-integrals.

Further inspection of the renormalized two-loop result reveals that logarithms in the EFT -- specifically those originating from the sunset integral $S_\text{3d}$ in Eq.~\eqref{3dsunset2loop} -- combine to a single logarithm of the form $\ln q_3$. This cancellation was made possible by setting the effective couplings $G_3$ and $\lambda_3$ to their leading order in the resummed contribution in Eq.~\eqref{Veffressoft2}. Alternatively, one could retain the effective couplings unexpanded, thereby including formally higher-order terms in the result (see, e.g., Ref.~\cite{Laine:2006cp} for an application in QCD). In either approach, the appearance of $\ln q_3$ reflects the cancellation of the 3d renormalization scale $\Lbar_\text{3d}$ between the naive and resummed soft contributions, yielding a logarithm of a scale ratio -- a hallmark of systems involving multiple scales.

To conclude this section, we stress that the IR divergence cancellation observed in the bosonic $\mathcal{S}$ sum-integral amounts to the simplest non-trivial scenario. At the three-loop level, such cancellation mechanism becomes significantly more intricate, as it requires analyzing specific kinematic limits of subgraphs within the full three-loop structure.

\section{Three loops and beyond}
\label{sec:threeloop}

With the aim of extracting accurate predictions for strong transitions in  BSM models,
the typically slow convergence of series expansions -- a generic feature of thermal gauge theories -- necessitates the inclusion of increasingly higher-order perturbative corrections. To this end, we proceed by outlining a general strategy that enables the systematic computation of the 4d resummed thermal effective potential within our framework, extending to three loops and beyond.

Given that computations in the 3d EFT amount to ordinary massive vacuum integrals -- a task amenable to well-developed tools at high orders (see, e.g., Refs.~\cite{Rajantie:1996np,Kajantie:2003ax}) -- what remains to be addressed is the evaluation of generic multi-loop vacuum sum-integrals. While the three-loop massless case is well understood (see Ref.~\cite{Schroder:2012hm} for a summary), the corresponding scenario involving massive propagators remains -- to the best of our knowledge -- entirely unexplored. To address this challenging task numerically, we develop a strategy centered on -- though not limited to -- the \texttt{hotLTD} technique.

In order to make generic massive sum-integrals amenable to numerical evaluation, divergences of both ultraviolet and infrared origin must be properly subtracted. In the former case, the problem is largely solved by the well-established $R$-operation in zero-$T$ vacuum QFT \cite{Bogoliubov:1957gp, Hepp:1966eg, Zimmermann:1969jj}, enabling a highly automatable construction of local UV subtraction terms \cite{Karkkainen:2025nkz} (see Appendix~\ref{app:hotLTD-example}). In contrast, the IR divergences are thermal in origin, arising from massless Matsubara zero modes propagating in loop diagrams. In this case, the required subtraction terms are conveniently embedded within the diagrammatic expansion of the EFT -- specifically in the object $V^\text{soft,naive}_\text{eff}$, an instance of which has already appeared in our evaluation of the fully bosonic $\mathcal{S}$ sum-integral in Eq.~\eqref{app:sunsetresults}. In this case, $S_\text{3d}$ is simply the fully zero-mode component of $\mathcal{S}$, acting as a local counterterm that smoothens out the otherwise singular infrared behavior as the boson mass approaches zero.

At the three-loop level, a new class of IR-sensitive sum-integral structures arises, encompassing graphs with fermion loops as well as those exhibiting power-like infrared sensitivity. Such contributions involve heavy modes propagating within $n$-point subgraphs embedded in the full graph, effects which are encoded in the effective parametres $X_3$ in the EFT through the loop-induced corrections $\delta X$. Accordingly, the appropriate subtraction terms for this class of graphs originate from an expansion of lower-order resummed integrals to higher orders in $\delta X$.

To illustrate how this mechanism plays out in practice, we consider a specific three-loop contribution to the effective potential in the scalar-Yukawa model. At this order, we find altogether 29 distinct three-loop sum-integrals in need of evaluation, of which 15 are non-factorized structures: these numbers result from applying the canonicalization algorithm introduced in Ref.~\cite{Navarrete:2024ruu} to our three-loop expressions, allowing to reduce the number of independent sum-integrals in need of evaluation. 
Including all prefactors resulting from the Feynman rules, a non-trivial three-loop contribution is
\begin{equation}
    \label{threeloopexample}
    \mathcal{M}= 4y^3G M_\psi \SumInt_{P}\frac{1}{P^2+M^2_\phi}\Pi_\text{b}(P,M_\phi)\Pi_\text{f}(P,M_\psi),
\end{equation}
where $\Pi_\text{b}$ and $\Pi_\text{f}$ are the bosonic and fermionic one-loop scalar two-point functions:
\begin{align}
    \Pi_\text{b}(P,M_\phi)&\equiv \SumInt_{Q} \frac{1}{(P^2+M^2_\phi)[(P-Q)^2+M^2_\phi]}, \label{scalar2pointbos} \\
    \Pi_\text{f}(P,M_\psi)&\equiv \SumInt_{\widetilde{Q}} \frac{1}{(P^2+M^2_\psi)[(P-Q)^2+M^2_\psi]}. \label{scalar2pointfer}
\end{align}
In order to investigate the infrared behavior of the sum-integral in $\mathcal{M}$ -- originating from the integration over soft momenta $P=(0,\mathbf{p})$ in the limit $|\mathbf{p}|\equiv p\ll \pi T$ -- we momentarily set $M^2_\phi=0$ and examine the behavior of the $\Pi_\text{b}$ and $\Pi_\text{f}$ two-point subgraphs at small external momentum. In the bosonic case, dimensional analysis dictates that \cite{Gynther:2007bw}
\begin{align}
    \Pi_\text{b}(p^0=0,\mathbf{p})\sim \frac{T}{p}+O(\varepsilon),
\end{align}
while the presence of heavy masses in all fermionic modes within $\Pi_\text{f}$ justifies setting $P=0$ directly, thereby reducing this function to a simple one-loop tadpole sum-integral
\begin{align}
    \Pi_\text{f}(P=0,M_\psi)=\SumInt_{\widetilde{Q}}\frac{1}{(Q^2+M_\psi^2)^2}. \label{softPif}
\end{align}
From this power-counting analysis it follows that $\mathcal{M}$ diverges logarithmically in the infrared, requiring a suitable counterterm to render the effective potential physically meaningful for small values of the bosonic mass.
\begin{figure}
    \centering
    \includegraphics[width=0.8\linewidth]{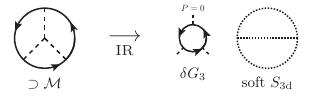}
    \caption{Infrared structure of the three-loop Mercedes diagram giving rise to our example sum-integral $\mathcal{M}$. In the soft limit $p\ll\pi T$, the fermion loop effectively decouples from the diagram, yielding a heavy-mode correction $\delta G_3$ to the 3d cubic coupling $G_3$. Meanwhile, the remaining zero-mode components reduce to the IR-sensitive sunset integral $S_\text{3d}$ from Eq.~\eqref{3dsunset2loop}, in the regime where the loop momenta are soft.}
    \label{fig:IRmercedes}
\end{figure}

As it turns out, $\mathcal{M}$ appears as a contribution to the three-loop ``Mercedes'' diagram depicted in Fig.~\ref{fig:IRmercedes}. Here, a graphical illustration for the infrared structure of this particular diagram serves to clarify the interpretation of $\Pi_\text{f}$, as given in Eq.~\eqref{softPif} in the soft limit, as a fermionic loop correction to the cubic coupling $\delta G_3$ within the EFT. Meanwhile, the remaining soft $P$ integration involving the $\Pi_\text{b}$ function amounts to the 3d sunset integral $S_\text{3d}$, exhibiting the logarithmic IR divergence anticipated for $\mathcal{M}$. In view of these considerations, the appropriate counterterm for $\mathcal{M}$ can be attributed to a term appearing in the unresummed expansion of the two-loop effective potential in the EFT -- namely, from the $V^\text{soft,naive}_\text{eff}$ object -- upon expanding $G_3$ to next-to-leading order. This specific contribution can be shown to be
\begin{equation}
\begin{split}
\label{threeloopexamplecounterterm}  V^{(3)\text{soft,naive}}_\text{eff}&\rvert_{\mathcal{M}}=\frac{1}{6}G \times \delta G_3\rvert_{\mathcal{M}} \times S_\text{3d}(M_\phi) \\
    &=4y^3 GM_\psi S_\text{3d}(M_\phi) \SumInt_{\widetilde{Q}}\frac{1}{(Q^2+M_\psi^2)^2},
\end{split}
\end{equation}
where the specific fermionic contribution to $\delta G_3$ required for our example is presented in more detail in Appendix~\ref{app:effectivecubiccoupling}. Importantly, $\delta G_3$ must be kept in unintegrated form in order to remove the singular IR behavior of $\mathcal{M}$ at the integrand level.

\begin{figure}
    \centering
    \includegraphics[width=1.0\linewidth]{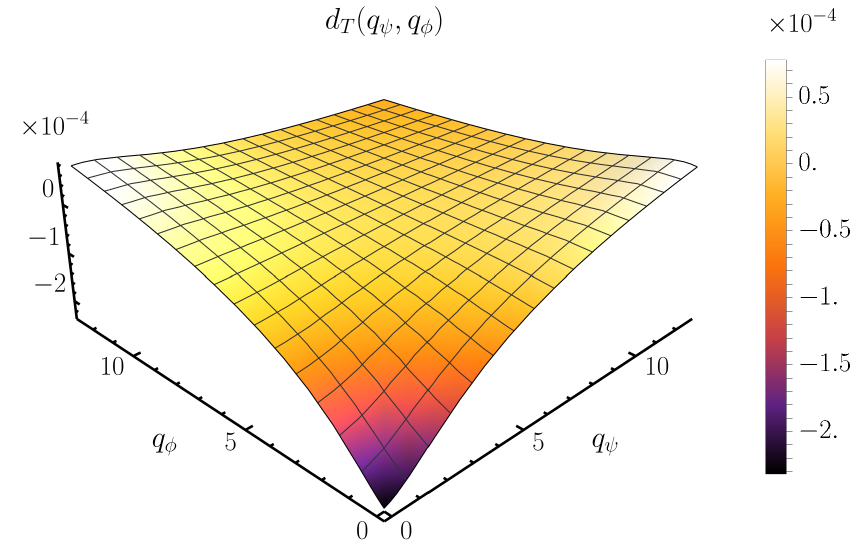}
    \caption{
    Three-loop thermal function defined in Eq.~\eqref{threeloopexampleresult}, obtained via numerical integration using \texttt{hotLTD} for mass-to-temperature ratios $q_{\psi,\phi}\in[0,4\pi]$. The resulting surface exhibits smooth behavior across the entire domain and approaches the $q_\phi\to0$ limit continously -- a regime where thermal resummations are essential. Numerical uncertainties are negligible and therefore not visible in the plot.}
    \label{fig:3loopexample}
\end{figure}

Collecting the expressions for Eqs.~\eqref{threeloopexample} and \eqref{threeloopexamplecounterterm}, the corresponding locally IR-safe combination contributing to the three-loop bare effective potential is
\begin{equation}
\begin{split}
\label{threeloopexampleresult}
    \Delta V^{(3) \text{hard}}_\text{eff}\rvert_{\mathcal{M}} &= \mathcal{M}-V^{(3)\text{soft,naive}}_\text{eff}\rvert_{\mathcal{M}} \\
    &= 4y^3GM_\psi \left( d_\text{UV}+T^2d_T\right).
\end{split}
\end{equation}
Here, $d_\text{UV}$ is a UV-divergent function of the masses and temperature resulting from the $R$-operation subtraction procedure, which includes the three-loop vacuum CW contribution and terms mixing lower-loop thermal and divergent vacuum components. Meanwhile, $d_T=d_T(q_\psi,q_\phi)$ is a dimensionless finite thermal function of the mass-temperature ratios, achieving a smooth interpolation to arbitrarily small values of the masses. Using \texttt{hotLTD}, we determine this function with high numerical accuracy across a wide range of masses, spanning from vanishing values up to heavy masses of order $4\pi T$ (i.e., $q_{\phi,\psi}\sim 4\pi$) \cite{ZenodoData}. The outcome of our numerical integration is presented in Fig.~\ref{fig:3loopexample} as a two-dimensional surface, clearly illustrating the expected smooth dependence on the bosonic mass parameter.

We now provide a comment on the remaining function $d_\text{UV}$ in Eq.~\eqref{threeloopexampleresult}. Given that determining the constant piece of $d_\text{UV}$ in dimensional regularization ($d=3-2\varepsilon$ spatial dimensions) requires knowledge of two-loop thermal contributions to linear order in $\varepsilon$ -- a formidable task lying beyond the scope of this work -- one could envision postponing this computation until the renormalization procedure has already been performed. In this case, poles in $\varepsilon$ multiplying thermal pieces cancel to all orders in the regulator against lower-loop contributions in the full effective potential, eliminating altogether the need to carry out such a challenging computation. This approach was previously employed in our evaluation of the two-loop effective potential, wherein one-loop thermal components multiplying poles were left unintegrated (see Eq.~\eqref{app:sunsetresults}). Ultimately, the remaining renormalized thermal contributions are to be determined following the strategy outlined earlier in the context of our two-loop computation.

To conclude this section, we turn to describe a procedure that covers the general case, concentrating on the identification of the appropriate IR counterterms that allow the construction of well defined hard contributions to the effective potential, $\Delta V^\text{hard}_\text{eff}$ -- an aspect of particular relevance in gauge theories.

\begin{enumerate}
    \item Generate all Feynman diagrams contributing to $V^\text{naive}_\text{eff}$ in the full theory, and cast the resulting sum-integrals into a standarized propagator basis -- for example, using the canonicalization procedure outlined in Ref.~\cite{Navarrete:2024ruu}).
    \item Repeat the procedure for $V^\text{res,soft}_\text{eff}$ and the effective parameters $X_3$ using the appropriate 3d effective theory describing the soft physics.
    \item By writing $X_3=X+\delta X_3$, generate the IR counterterms in $V^\text{naive,soft}_\text{eff}$ by expanding in $\delta X_3$ to the corresponding order in the 4d-theory couplings.
    \item To construct $\Delta V^\text{hard}_\text{eff}$, the resulting expressions are matched at the integrand level to IR-sensitive sum-integrals -- for instance, by employing pattern matching techniques \cite{Vermaseren:2000nd}.
\end{enumerate}

Together with the systematic subtraction of UV divergences via the $R$-operation, every conceivable contribution to the hard effective potential $\Delta V^\text{hard}_\text{eff}$ can be cast into the canonical form of Eq.~\eqref{threeloopexampleresult}. Combined with the corresponding resummed terms in $V^\text{soft,res}_\text{eff}$, this procedure enables a systematic determination of the renormalized effective potential at three-loop order and beyond. Naturally, our perturbative program ultimately encounters fundamental limitations at the four-loop level in non-Abelian gauge theories, where non-perturbative effects become unavoidable. Resolving this formidable challenge likely requires augmenting perturbative results with input from dedicated lattice computations.

\begin{figure*}[ht]
  \begin{centering}
    \includegraphics[width=1.0\columnwidth]{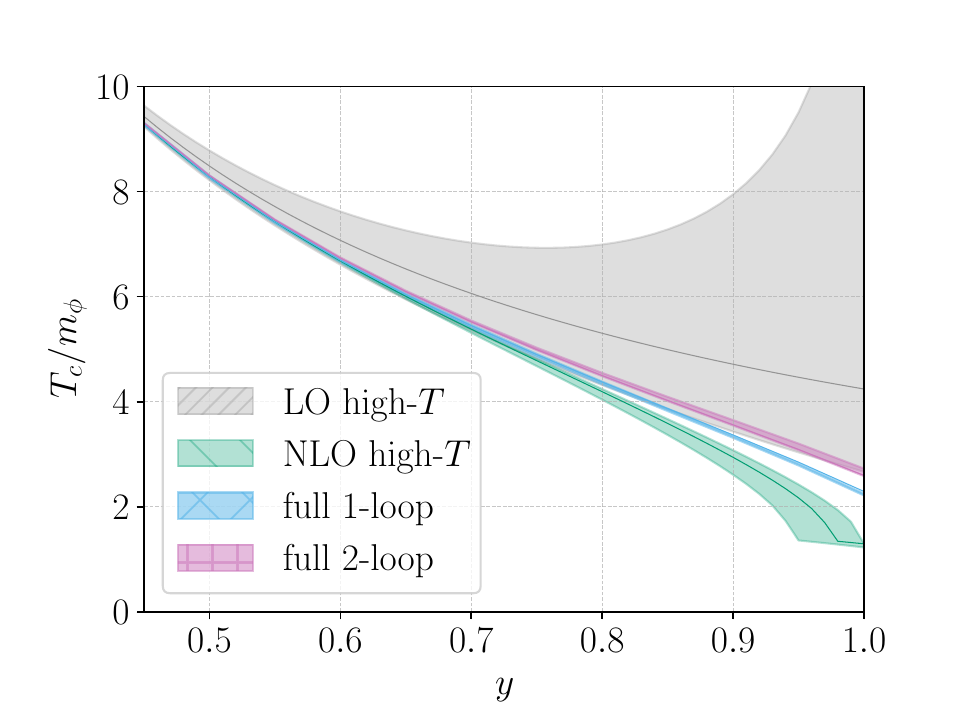}
    \includegraphics[width=0.92\columnwidth]{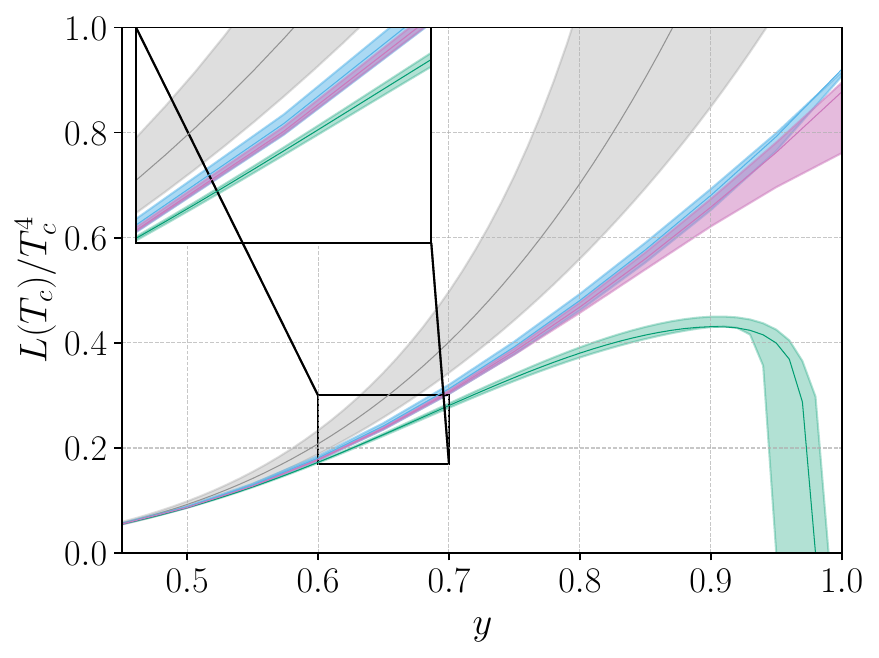} 
  \end{centering}
  \caption{
Critical temperature (left) and latent heat (right) as function of $y$ in the benchmark point of Eq.~\eqref{eq:BM} in the high-$T$ expansion at leading (gray) and next-to-leading (green) orders, together with full results without the high-$T$ expansion at one (blue) and two loops (purple). For each approximation, band depicts the variation with respect to renormalization scale, as explained in the main text. Crucially, at large $y$, the full calculation without high-$T$ expansions results in significantly stronger transitions compared to the 3d EFT approach at NLO in the high-$T$ approximation, and furthermore avoids the total breakdown plaguing the latter.
}
  \label{fig:results}
\end{figure*}

We now proceed to a detailed analysis of the phase transition parameters as derived from our full one- and two-loop computations in Eqs.~\eqref{eq:Veffren1loop} and \eqref{eq:veffren2loop}.

\section{Thermodynamics}
\label{sec:thermo}

Equilibrium thermodynamic properties can be determined from the 
difference of the pressure $p$ (minus free energy density) between the two phases. The pressure associated with a given phase is determined by the effective potential via $p = - V_\text{eff}(\varphi_{\text{min}})$, where $\varphi_{\text{min}}$ denotes the field value at the minimum corresponding to that phase. With this quantity at hand, two key thermal parameters characterizing the phase transition can be determined. First, the critical temperature $T_c$ at which the transition occurs is solved from the condition $\Delta V_\text{eff}(T_c) = 0$, where $\Delta$ refers to the difference between the high- and low-temperature phases. Second, the latent heat released during the transition, which quantifies the strength of the process, is given by $L(T) = T \Delta dp/dT$. We note in passing that, while $\varphi_\text{min}$ lacks any physical meaning -- being subject to additive or multiplicative field redefinitions -- the difference $\Delta V_\text{eff}$ remains invariant under such transformations. As a result, both the critical temperature $T_c$ and the latent heat $L$ are well-defined physical observables.

Following Ref.~\cite{Chala:2024xll}, for a numerical illustration we choose 
\begin{align}
\label{eq:BM}
\sigma &= 0, \nonumber \\
m^2_\phi &= 20000 \; \text{GeV}^2 , \nonumber \\
m_\psi &= 0, \\
\lambda &= 0.24, \nonumber \\
g &= -240 \; \text{GeV}, \nonumber   
\end{align}
and vary $y \in [0.45,1]$. Given that the theory we consider is a toy model, no sensible physical observables can enable relations among $\overline{\text{MS}}$ parameters and physical ones at some zero-$T$ scale. In Ref.~\cite{Gould:2023jbz}, these parameters are fixed at an initial $\overline{\text{MS}}$ scale $\Lbar_0 = m_\phi$, while in Ref.~\cite{Chala:2024xll} the thermal scale $\Lbar_0 = \pi T e^{-\gamma_{\rm{E}}}$ is chosen instead (here $\gamma_{\rm{E}}$ denotes the Euler–Mascheroni constant). 

In this work, we define the $\overline{\text{MS}}$ parameters of Eq.~\eqref{eq:BM} at an initial scale $\Lbar_0 = \pi T_c^{\text{LO}}$, corresponding to the leading-order  critical temperature as derived analytically in a high-$T$ expansion. While we emphasize that this choice is not motivated by physical considerations, it facilitates comparisons across different perturbative orders. In addition, it enables an assessment of the accuracy of the high-$T$ expansion for strong transitions and facilitates comparisons against results derived in Ref.~\cite{Chala:2024xll} within the high-$T$ approximation. Finally, to evaluate the sensitivity of our perturbative results to higher-order corrections, the thermal renormalization scale is varied within the range $\Lbar \in [0.5\pi T_c^{\text{LO}}, 2\pi T_c^{\text{LO}}]$ together with the corresponding renormalization-group running of the parameters. To this end, the required $\beta$-functions are listed in Appendix~\ref{sec:betas}.

Before proceeding, we comment on a subtlety specific to this toy model. If one identifies the initial scale $\Lbar_0$ with a more physically motivated zero-$T$ scale, sufficiently large values of $y$ drive the scalar self-coupling $\lambda$ to turn negative already at the thermal scale, while inducing large logarithms that undermine perturbative control. As a result, the zero-$T$ Coleman-Weinberg potential may become unstable at relatively small field values, preventing a meaningful analysis within this model. In view of these observations, and given that our primary goal is to test the reliability of the high-$T$ expansion for strong transitions at large $y$ values, we do not pursue a realistic physical treatment here. A more complete picture would require, for instance, renormalization-group improvement together with a field-dependent renormalization scale to ensure the stability of the zero-$T$ potential at large field values (see, e.g., Ref. \cite{Kierkla:2023von}).

Turning to the thermodynamics, Fig.~\ref{fig:results} displays our results for $T_c$ and the $L(T_c)/T_c^4$ as functions of $y$. The blue and purple bands represent our main results, obtained from the full one- and two-loop effective potentials (see Eqs.~\eqref{eq:Veffren1loop} and \eqref{eq:veffren2loop}), without relying on the high-$T$ expansion. For comparison, the corresponding LO and NLO results derived within the high-$T$ expansion are shown in gray and green, respectively. Details of the LO and NLO definitions can be found in Appendix~\ref{sec:3D-EFT}. As $y$ increases, the critical temperature decreases while the strength of the transition correspondingly grows. In each case, the theoretical uncertainty associated with renormalization-scale variation is indicated by a shaded band around the central value $\Lbar_0 = \pi T_c^{\text{LO}}$.

Let us now dissect the various approximations in more detail. For the leading order results within the high-$T$ expansion, the broad uncertainty bands reflect the significant scale ambiguity arising from missing higher-order logarithmic terms, which would otherwise compensate for the scale dependence of the running parameters in the tree-level potential. While this running is formally a higher-order effect, the pronounced scale sensitivity at large $y$ values signals that the LO computation becomes unreliable in this regime. The necessity of higher-order corrections in accurately describing strong transitions becomes apparent upon observing that such transitions occur in regions of parameter space where $\lambda / y \ll 1$. This hierarchy implies a large mass separation between the transitioning scalar and the fermion responsible for the first-order phase transition, suggesting in addition that fermionic loop corrections influence the properties of the transition more strongly than loop effects from the scalar sector. Within the high-$T$ expansion this feature is evident, as terms polynomial in $y$ substantially modify the effective scalar self-coupling (see Eq.~\eqref{eq:lam3} in Appendix~\ref{sec:3D-EFT}) and the zero-$T$ running of $\lambda$ from Eq.~\eqref{eq:lam-run}. This highlights the critical role of loop corrections from the transition-inducing field in the regime of strong transitions, and the necessity of incorporating all NLO contributions -- including the running of parameters. Such effects were not accounted for in Ref.~\cite{Chala:2024xll} for the same benchmark point given in Eq.~\eqref{eq:BM}.

This intrinsic perturbative uncertainty is significantly reduced by including higher-order corrections. Within the high-$T$ expansion, this is achieved via the higher-order loop computation of the 3d effective potential together with NLO dimensional reduction, which amounts to determining two-loop thermal masses and one-loop thermal couplings (see Appendix~\ref{sec:3D-EFT} for details). These results are shown in a green hue, demonstrating a clear improvement due to a much narrower uncertainty band.

Notably, within the 3d EFT computation, phase transitions terminate entirely around $y \approx 0.97$, as the parent theory no longer maps into a regime of the 3d EFT that allows a phase transition. Specifically, the broken minimum of the 3d effective potential never becomes degenerate with the high-$T$ minimum. Instead, as the temperature decreases, the potential in the broken phase initially drops but begins to rise again beyond a certain point; this is visible in the left panel of Fig.~\ref{fig:results}, where the green curves abruptly terminate (see Appendix~\ref{sec:3D-EFT} for details). We emphasize that this somewhat peculiar behavior is a distinctive feature of the toy model under consideration, arising from specific choice of parameter values and the renormalization-group running procedure. Nevertheless, this behavior serves as a clear indication of the breakdown of the high-$T$ expansion, as it does not arise in the full effective potential evaluated without the high-$T$ approximation.

Indeed, our full results without the high-$T$ expansion depict a great improvement for the strongest transitions at large $y$: already at one-loop order (blue), the result exhibits minimal sensitivity to variations of the renormalization scale -- virtually negligible for $T_c$ and very small for the latent heat -- even at the largest $y$ values, where the transition is extremely strong.
Adding the two-loop contributions (purple) provides minor corrections, except for the very strongest transitions at large $y$ values due to the overall convergence becoming worse, as signaled by the increasing sensitivity to the renormalization scale. On the other hand, at smaller $y$ values such sensitivity is properly reduced compared to the mere one-loop result. To facilitate this observation, we have zoomed in to a relevant region with $y \in [0.6,0.7]$ in the right panel of Fig.~\ref{fig:results}.

Overall, at large $y$ values the full calculation without the high-$T$ expansion results in significantly smaller critical temperatures and larger latent heat compared to the 3d EFT approach. At small $y$ values, where the transition is very weak, all approaches yield consistent results, as expected, and the validity of the high-$T$ approximation is well established. The visible deviation between the blue/purple and green curves for $y \gtrsim 0.7$ signals the decreasing reliability of the high-$T$ expansion. Moreover, the rapidly widening green band in $T_c$ highlights the onset of a breakdown in this approximation.

Although not visualized in Fig.~\ref{fig:results}, we have explicitly verified that the discrepancy between the approaches with and without the high-$T$ expansion stems almost entirely from the differing treatment of the fermionic contributions responsible for driving the transition. Around $T_c$, the transitioning scalar remains much lighter than the fermion, allowing for a reliable high-$T$ expansion of scalar contributions. Conversely, the background-field-dependent fermion mass grows significantly with the transition strength, typically reaching $O(\pi T)$ values for strong transitions. Notably, the primary difference with the high-$T$-expanded results originates from the fermionic thermal contributions to the effective potential itself, rather than from fermionic corrections to the soft thermal mass of the scalar. We expect these features to be broadly general: near the phase transition, the transitioning fields typically remain light, while the transition-inducing fields -- whether fermions or bosons in generic models -- become increasingly heavy in the low-$T$ phase in strong transitions. As a result, the high-$T$ expansion breaks down for loops involving these large, field-dependent masses. See Appendix~\ref{sec:portal-sector} for further discussions on gauge-Higgs theories featuring strong first-order phase transitions.

Finally, to conclude our numerical investigations in the scalar-Yukawa model, we have verified that varying the couplings $\lambda$ and $y$ yields qualitatively similar behaviors. A more comprehensive exploration of the parameter space would require varying the remaining parameters -- $m_\phi$, $m_\psi$, and $g$ -- along the lines of the scan performed in Ref.~\cite{Gould:2024chm}.

\section{Discussion}
\label{sec:discussion}

In this work, we have developed a new perturbative framework to compute equilibrium thermodynamic properties of cosmological phase transitions to high loop orders using the full four-dimensional resummed thermal effective potential, without resorting to the standard high-$T$ expansions. As a concrete demonstration of our framework, we explored strong phase transitions in a scalar–Yukawa model, presenting a complete two-loop calculation together with a novel non-trivial three-loop example that pushes the frontier of thermal field theory computations.

The main conceptual advantage of our framework is that it provides a unified description of hard and soft scales within a single expression, valid order by order in perturbation theory. This in turn clarifies several conceptual and practical issues that have hindered progress in perturbative studies of thermal phase transitions. In particular, the local (i.e., integrand-level) cancellation of vacuum and thermal divergences is achieved in a transparent way through a systematic construction of UV and IR counterterms, holding the potential to be fully automated to high loop orders. This allows to render thermal sum-integrals in a form amenable to numerical evaluation, for which we have employed a finite-temperature generalization of the Loop-Tree Duality \cite{Capatti}. Furthermore, our framework sheds new light on the problem of imaginary components in the effective potential originating from tachyonic masses, a discussion largely neglected in the literature (however, see \cite{Croon:2020cgk,Lofgren:2023sep,Athron:2023xlk} and references therein). Concretely, we have argued that such issues are specific to the resummed soft sector, arising only near the potential barrier far from the physical minima and thus leaving thermodynamic properties unaffected.

Our framework not only provides a precise way to test the accuracy of high‑$T$ expansions for moderately strong transitions -- those still accessible by the 3d EFT -- but further extends perturbative calculations into the regime of the strongest transitions, where such approximations fail. In general, strong phase transitions require a significant mass hierarchy between the transitioning scalar field and a heavier transition-inducing field in the low-$T$ phase, which leads to a breakdown of the high-$T$ expansion. Due to this generic feature, our new framework opens up wide opportunities for accurately studying the thermodynamics of models with extremely strong phase transitions, particularly gauge-Higgs theories relevant for the EW phase transition.

It was demonstrated in Ref.~\cite{Gould:2019qek} that a first-order phase transition in \textit{any} BSM scenario, that can be reliably described within the high-$T$ expansion in the minimal 3d EFT for the EW theory, cannot produce gravitational-wave signals strong enough to be observed by LISA or any planned detectors. This ``no-go'' conclusion holds provided no new light transitioning fields beyond the SM Higgs scalar are involved. To avoid this conclusion within the high-$T$ expansion, one must either introduce new light fields that enable a multi-step transition \cite{Niemi:2020hto,Friedrich:2022cak,Gould:2023ovu}, or add corrections from higher-dimensional thermal operators that emerge from a slowly converging perturbative expansion at high temperatures.

Recent studies have aimed at exploring in increasing detail the effects of higher-dimensional operators in thermal phase transitions \cite{Chala:2024xll,Niemi:2024vzw,Bernardo:2025vkz,Chala:2025aiz}. While these works have so far remained inconclusive regarding the overall validity of the high-$T$ expansion in the regime of sufficiently strong transitions, they provide concrete evidence that such operators have sizable effects on the thermodynamics of the transition. This challenges not only the widespread use of low-order truncated high-$T$ expansions in perturbative calculations, but also casts doubt on otherwise robust non-perturbative lattice studies of thermodynamics and bubble nucleation rates -- where incorporating such operators is far from straightforward.

By contrast, abandoning the high‑$T$ expansion altogether would bypass the no‑go conclusion of Ref.~\cite{Gould:2019qek}. This suggests that any attempt to obtain an observable gravitational‑wave signal from a single‑step EW phase transition may require avoiding the high‑$T$ expansion for the transition‑inducing fields (see Appendix~\ref{sec:portal-sector} for further discussion). In its present form -- focused on the equilibrium thermodynamics of such strong transitions -- our newly developed framework offers a concrete tool for scrutinizing these issues.

Finally, we can envision clear future directions for applying and building upon our framework. The equilibrium properties of several concrete BSM theories could be investigated without relying on the high‑$T$ expansion (see e.g., Ref.~\cite{Laine:2017hdk}), enabling unprecedented precision and allowing one to scrutinize the reliability of current state‑of‑the‑art lattice simulations within 3d EFTs \cite{Kainulainen:2019kyp,Niemi:2020hto,Niemi:2024axp}. Alternatively, results obtained within our framework could be directly contrasted to full 4d lattice simulations (see Refs.~\cite{Csikor:1996sp,Laine:1996nz,Csikor:1998eu,Laine:1999rv} and references therein), offering a means to assess the applicability and accuracy of perturbation theory more broadly. Importantly, our computational techniques could also be extended to perturbative evaluations of the effective action and bubble nucleation rate \cite{Moore:2000jw,Ekstedt:2021kyx,Gould:2021ccf,Gould:2022ran,Gould:2024chm}, as well as the sphaleron rate \cite{Moore:1998swa,Li:2025kyo,Annala:2025aci}, both of which are relevant for preserving any baryon asymmetry generated during a strong EW phase transition. This would represent a crucial step toward obtaining reliable predictions relevant for LISA‑generation gravitational-wave measurements, and for singling out viable models of EW baryogenesis -- specially if studies within our framework reveal that strong transitions in SM extensions suffer from a breakdown of the high‑$T$ expansion. The conclusion drawn from such studies may play a key role in cracking open questions in the field of cosmology, while reshaping our understanding of thermal phase transitions in particle physics.

\section*{Acknowledgments}

We thank 
F. Bernardo,
M. Chala,
D. Curtin,
A. Ekstedt,
O. Gould,
J. Hirvonen,
M. Kierkla,
P. Klose,
A. Kurkela,
J.  L{\"o}fgren,
L. Niemi,
K. Rummukainen,
P. Schicho,
Y. Schr{\"o}der,
B. {\'S}wie{\.z}ewska,
J. van de Vis,
and
C. Xie
for discussions. In particular, we thank K. Kajantie and A. Vuorinen for helpful comments and suggestions. 
Our work has been supported by the Research Council of Finland grants 347499,
353772, 354533, and 354572, as well as by the European Union (ERC, CoCoS, 101142449).

\section*{Data availability}

All thermal functions obtained numerically via the \texttt{hotLTD} algorithm are archived and publicly available in Zenodo \cite{ZenodoData}.

\appendix

\section{Sum-integrals}
\label{app:details}

Throughout our computation, Euclidean propagators defined in terms of four-momenta $P = (p_0, \pt)$ are employed. Here, the Matsubara frequencies read $p_0 = 2n\pi T$ for bosons and $p_0 = (2n+1)\pi T$ for fermions, where $n \in \mathbb{Z}$. Also, we make use of the notation $\vert \pt \vert \equiv p$ for three-vector magnitudes.

The sum-integral symbol for bosonic momenta $P$ is defined as 
\begin{equation}
\SumInt_{P} f(P) \equiv T\sum_{n\in\mathbb{Z}}\int_{\pt} f(2\pi nT,\mathbf{p}), 
\end{equation}
while momenta with fermionic frequencies are represented by $\widetilde{P}$ instead. The $d$-dimensional spatial part of the integration measure is given in the $\overline{\text{MS}}$ renormalization scheme by
\begin{equation}
\label{app:3dmeasure}
\int_{\pt}  \equiv \left (\frac{e^{\gamma_{\text{E}}}\Lbar^2}{4\pi} \right )^{\frac{3-d}{2}}\int \frac{\mathrm{d}^d\mathbf{p}}{(2\pi)^d}, \end{equation}
where $\Lbar$ is the renormalization scale (in the EFT, we evaluate integrals with a 3d scale $\overline{\Lambda}_\text{3d}$ instead) and $\gamma_{\text{E}}$ is the Euler-Mascheroni constant, while integrals are regulated by employing dimensional regularization in $d=3-2\varepsilon$. We also make use of the notation in Eq.~\eqref{app:3dmeasure} with $\int_P$ instead to denote $D=d+1$-dimensional vacuum ($T=0$) integrals.

In the following, we present our results for all required sum-integrals and integrals up to the constant term in dimensional regularization, i.e. to $O(\varepsilon^0)$.

The basic ``one-loop'' massive sum-integral contributing to the leading-order thermal corrections is (here $m^2\geq0$)
\begin{align}
    \SumInt_{P} \ln\left(P^2+m^2\right) = \int_P \ln\left(P^2+m^2\right)+ \frac{T^4}{\pi^2}h(m/T),
\end{align}
where the vacuum part reads
\begin{align}
    \int_P \ln\left(P^2+m^2\right)=\frac{-m^4}{2(4\pi)^2}\left(\frac{1}{\varepsilon}+\ln\frac{\Lbar^2}{m^2}+\frac{3}{2}\right), \label{appeq:logvacuum}
\end{align}
and the thermal contribution can be expressed in terms of the numerical integral
\begin{align}
    h(m/T) \equiv \int_0^\infty \text{d}x\, x^2\ln(1-e^{-\sqrt{x^2+(m/T)^2}}). \label{appeq:hfunction}
\end{align}
In the case of $m^2<0$, the $h$ function may be combined at integrand level with the vacuum and zero-mode-subtraction pieces in the full effective potential. The resulting expression is real and finite in the entire range $m^2>-(2\pi T)^2$, and can be written as (a prime denotes that the zero-mode is removed)
\begin{widetext}
\begin{equation}
\begin{split}
    \SumIntprime_{P} \ln(P^2+m^2)= \frac{-m^4}{2(4\pi)^2}\left(\frac{1}{\varepsilon}+\ln\frac{\Lbar^2}{|m|^2}+\frac{3}{2}\right) + \frac{T^4}{2\pi^2} \int_0^\infty \text{d}x\,x^2\text{Re}\ln\left[\frac{1-e^{-\sqrt{x^2+(m/T)^2}}}{x^2+(m/T)^2}\right].
\end{split}
\end{equation}
\end{widetext}
In our computation, massive one-loop bosonic and fermionic tadpole sum-integrals -- possibly with higher propagator powers -- are required. These are defined as:
\begin{equation}
\begin{split}
    \mathcal{I}_\nu(m,T) & \equiv \SumInt_{P} \frac{1}{(P^2+m^2)^\nu}, \\ \widetilde{\mathcal{I}}_\nu(m,T)& \equiv \SumInt_{\widetilde{P}} \frac{1}{(P^2+m^2)^\nu}. \label{deftadpoles}
\end{split}
\end{equation}
In this paper, we need all cases up to $\nu=3$; the fermionic tadpoles are straightforward to obtain in terms of the bosonic one via the relation
\begin{align}
    \widetilde{\mathcal{I}}_\nu(m,T)=2^{2\nu-d}\mathcal{I}_\nu(2m,T)-\mathcal{I}_\nu(m,T).
\end{align}
In some parts of the text, we make use of a notation that distinguishes vacuum and thermal parts,
\begin{equation}
    \mathcal{I}_\nu(m,T)=\mathcal{I}^\text{V}_\nu(m)+\mathcal{I}^T_\nu(m,T), \label{appeq:tadpolesplit}
\end{equation}
where $\mathcal{I}^\text{V}(m)\equiv \lim_{T\to0}\mathcal{I}(m,T)$. Note that we suppress the $\nu$ index for the case of $\nu=1$,                                                       i.e. $\mathcal{I}\equiv\mathcal{I}_1$.

The result for $\nu=1$ is well known (see, e.g., Ref.~\cite{Laine:2016hma}). Making use of the Bose-Einstein distribution function $n_\text{B}(x)\equiv(e^x-1)^{-1}$ and denoting the normalized energy by $E\equiv[x^2+(m/T)^2]^{1/2}$, it reads (here $m^2\geq0$)
\begin{equation}
    \mathcal{I}=\frac{-m^2}{(4\pi)^2}\left( \frac{1}{\varepsilon}+\ln\frac{\Lbar^2}{m^2}+1 \right)+\frac{T^2}{2\pi^2}r(m/T), \label{appeq:tadpole1}
\end{equation}
where the thermal part is given by
\begin{align}
    r(m/T)\equiv \int_0^\infty \text{d}x\,x^2\frac{n_\text{B}(E)}{E}.
\end{align}
In the full effective potential and matching coefficients, the $\mathcal{I}_\nu$ tadpoles always appear subtracted with their zero-mode counterparts, enabling to extend their definition to $m^2>-(2\pi T)^2$. In our computation, we require all cases up to $\nu=3$, which we write with a prime symbol on top to indicate that only non-zero modes contribute. The resulting expressions can be written as follows (a prime in the distribution function denotes a derivative with respect to its argument):
\begin{widetext}
\begin{equation}
\begin{split}
\label{appeq:fulltadpoles}
    \mathcal{I}'&=\frac{-m^2}{(4\pi)^2}\left( \frac{1}{\varepsilon}+\ln\frac{\Lbar^2}{|m|^2}+1\right) + \frac{T^2}{2\pi^2} \int_0^\infty \text{d}x\,x^2\text{Re}\left[\frac{n_\text{B}(E)}{E}-\frac{1}{E^2}+\frac{1}{x^2}\right], \\
    \mathcal{I}'_2&=\frac{1}{(4\pi)^2}\left[ \frac{1}{\varepsilon}+2\ln\frac{\Lbar}{2\pi T}+2\ln (\pi e)\right] + \frac{1}{4\pi^2}\int_0^\infty \text{d}x\,x^2 \,\text{Re}\left[\frac{-n'_\text{B}(E)}{E^2} + \frac{n_\text{B}(E)}{E^3} -\frac{2}{E^4} +\frac{1}{2E^3}-\frac{1}{2x}\frac{1}{1+x^2}\right], \\
    \mathcal{I}'_3&=\frac{1}{T^2}\frac{1}{(4\pi)^2}\int_0^\infty \text{d}x\,x^2\,\text{Re}\left[ \frac{n''_\text{B}(E)}{E^3}-\frac{3n'_\text{B}(E)}{E^4}+\frac{3n_\text{B}(E)}{E^5}-\frac{8}{E^6}+\frac{3}{2E^5} \right].
\end{split}
\end{equation}
\end{widetext}
Turning to two loops, the required sum-integrals are
\begin{align}
    \mathcal{S}(m)&\equiv\SumInt_{PQ}\frac{1}{(P^2+m^2)(Q^2+m^2)(R^2+m^2)}, \label{app:bosonicsunset} \\
    \widetilde{\mathcal{S}}(m_1,m_2)&\equiv\SumInt_{\widetilde{P} \widetilde{Q}}\frac{1}{(P^2+m_1^2)(Q^2+m_1^2)(R^2+m_2^2)},\label{app:fermionicsunset}
\end{align}
where $R=P-Q$. Upon a suitable subtraction of divergences as described in Appendix~\ref{app:hotLTD-example}, these sum-integrals can be cast in the form
\begin{widetext}
\begin{equation}
\begin{split}
\label{app:sunsetresults}
    \mathcal{S}(m)&= S_\text{vac}(m)+S_\text{3d}(m)-\frac{1}{(4\pi)^2}\left\{T^2\left(\frac{1}{4\varepsilon}+\frac{1}{2}+\ln\frac{\Lbar_\text{3d}}{6\pi T} \right) -3\mathcal{I}^T(m,T)\left(\frac{1}{\varepsilon}+2\ln\frac{\Lbar}{2\pi T}\right)\right\}+T^2b_T(q)+O(\varepsilon), \\
    \widetilde{\mathcal{S}}(m_1,m_2)&=S_\text{vac}(m_1,m_2)+\frac{1}{(4\pi)^2}\left[ 2\widetilde{\mathcal{I}}^T(m_1,T)+\mathcal{I}^T(m_2,T) \right]\left(\frac{1}{\varepsilon}+2\ln\frac{\Lbar}{2\pi T}\right)+T^2f_T(q_1,q_2) +O(\varepsilon) .
\end{split}
\end{equation}
\end{widetext}
Here, $b_T$ and $f_T$ are finite two-loop bosonic and fermionic thermal functions that we compute numerically to high precision with \texttt{hotLTD} \cite{ZenodoData} ($b_T$ and $f_T$ can also be extracted from, e.g., \cite{Laine:2017hdk}). For each sum-integral in Eq.~\eqref{app:sunsetresults}, the term $S_\text{vac}$ denotes the temperature-independent contribution, while $S_\text{3d}$ is the fully zero-mode part of the bosonic $\mathcal{S}$. The latter integral exhibits a logarithmic IR divergence in the $m\to0$ limit; consequently, the function $b_T$ remains well behaved across the entire mass range. For our results above, we have made use of the one-loop thermal parts $\mathcal{I}^T$ and $\widetilde{\mathcal{I}}^T$ defined in Eq.~\eqref{appeq:tadpolesplit}, which can be extracted from Eq.~\eqref{appeq:tadpole1}. As described in the one-loop cases, the extension of these results to the $m^2<0$ regime can be carried out by subtracting the relevant thermal counterterms directly at the integrand level -- namely, the contributions in Eq.~\eqref{Veffnaivesoft2}, most notably the mass correction $\delta M^2_3$ given in Eq.~\eqref{appeq:deltaM3}.

Within the EFT, three 3d integrals are required to $O(\varepsilon^0)$ in dimensional regularization. Assuming $m^2\geq0$, the one-loop cases read
\begin{equation}
\begin{split}
\label{3dintegrals1loop}
    &\int_\mathbf{p}\ln(p^2+m^2)=  -\frac{1}{6\pi}(m^2)^{3/2}, \\
    I_{3d}&(m) \equiv T\int_{\pt} \frac{1}{p^2+m^2}=-\frac{mT}{4\pi},
\end{split}
\end{equation}
while the 2-loop 3d sunset integral is \cite{Farakos:1994kx}
\begin{equation}
\begin{split}
\label{3dsunset2loop}
    S_{3d}(m) &\equiv T^2\int_{\pt\qt} \frac{1}{(p^2+m^2)(q^2+m^2)(r^2+m^2)} \\
    &= \frac{T^2}{(4\pi)^2}\left(\frac{1}{4\varepsilon}+\ln\frac{\overline{\Lambda}_\text{3d}}{3m}+\frac{1}{2}\right),
\end{split}
\end{equation}
where we have denoted $\mathbf{r} = \mathbf{p}-\mathbf{q}$. Making use of these results in the $m^2<0$ regime may lead to the emergence of imaginary components near the potential barrier in the case of the resummed contributions, where the (perturbative) squared thermal mass is expected to be tachyonic.

The zero-temperature contribution to the two-loop effective potential displayed on the first line in Eq.~\eqref{eq:veffhard2} requires knowledge of the 4d vacuum sunset integral (the $S_\text{vac}$ from Eq.~\eqref{app:sunsetresults}), which we take from Ref.~\cite{Caffo:1998du}. The resulting expressions read:
\begin{widetext}
\begin{equation}
\begin{split}
\label{app:twoloopCW}
    B_1(x)&= \frac{1}{8}\ln^2(x^2)+\frac{1}{4}\ln(x^2)+\frac{1}{8}, \\
    B_2(x)&= \frac{1}{8}\ln^2(x^2)+\frac{1}{2}\ln(x^2) +\frac{5}{8}
    -\frac{1}{72}\left[ \psi_1\left(1/3\right)+\psi_1\left(1/6\right)-\frac{8\pi^2}{3}\right], \\
    F(x_1,x_2)&=-3\ln^2(x_1^2)-14\ln(x_1^2)-19 + z^2\left\{ 8w\text{Li}_2\left[1-\frac{z^2}{2}(1+w)\right]+2w\ln^2\left[-1+\frac{z^2}{2}(1+w)\right]+\frac{2\pi^2}{3}w\right. \\
    &\left.+3\ln^2(x_1^2)-6\ln(x_1^2)\ln(x_2^2)+2\ln(x_1^2)-10\ln(x_2^2)-7\right\} - z^4 \left\{ 2w\text{Li}_2\left[ 1-\frac{z^2}{2}(1+w) \right]\right. \\
    &\left.+\frac{w}{2}\ln^2\left[-1+\frac{z^2}{2}(1+w)\right]+\frac{\pi^2}{6}w+\frac{1}{2}\ln^2(x_1^2)-\ln(x_1^2)\ln(x_2^2)-2\ln(x_2^2)-\frac{5}{2} \right\}.
\end{split}
\end{equation}
\end{widetext}
Here, we define $z^2\equiv x_1^2/x_2^2$ and $w\equiv\sqrt{1-4z^{-2}}$, while $\psi_1(x)=d^2/dx^2 \ln\Gamma(x)$ is the trigamma function and Li$_2(x)$ is the dilogarithm \cite{Abram:1972book}. Note that, although the fermionic function $F$ involves complex terms for $z^2>0$, it is a real function. Furthermore, as an example of the asymptotic behavior of $F$ in the massless case, by setting $z=1$ (i.e., $x_1=x_2\equiv x$) and then taking $x^2\to0$, its leading behavior is $F(x^2)\approx-7\ln^2(x^2)$; in general, $F$ does not grow faster than $\ln^2(x^2)$. Finally, the extension of these functions to tachyonic masses $m^2<0$ is implemented within the full effective potential by simply replacing squared masses by their absolute values, in accordance with the discussion at the end of Sec.~\ref{sec:pop}.

\section{\texorpdfstring{Matching parameters without high-$T$ expansions}{Matching parameters without high-T expansions}}
\label{app:matchingparameters}

Perturbative matching of various correlators between the EFT and the full theory grants the matching parameters entering the 3d Lagrangian in Eq.~\eqref{3dlagrangian}. Lifting the high-$T$ expansion, here we derive the required one-loop effective mass $M^2_3$, and describe a specific one-loop contribution to $G_3$ needed for the three-loop example described in Sec.~\ref{sec:threeloop}. To this end, we follow Ref.~\cite{Kajantie:1995dw} closely.

\subsection{Effective mass parameter}
\label{app:effectivemass}

After enough resummations, the (inverse) renormalized static two-point function for the scalar field in the full theory can be written as
\begin{align}
    k^2+M^2_\phi(\varphi) +\Delta \Pi^\text{hard}_\text{R}(k^2)+\Pi^\text{soft,res}_\text{R}(k^2). \label{appeq:4dcorr}
\end{align}
Owing to $\Delta\Pi^\text{hard}_\text{R}$ capturing hard-scale effects only, it is well defined in the infrared and can hence be expanded in soft momenta $k^2\lesssim \lambda T^2$ up to $O(\lambda^2T^2)$. This yields
\begin{align}
    \Delta\Pi^\text{hard}(k^2)\approx\Delta\Pi^\text{hard}_\text{R}(0)+k^2\Delta\Pi^{'\text{hard}}_\text{R}(0),
\end{align}
where $\Delta\Pi^{'\text{hard}}(k^2)\equiv d/dk^2 \Delta \Pi^\text{hard}(k^2)$.

In the EFT, the analogous quantity for the light zero-mode bosonic field reads
\begin{align}
    k^2+M^2_3(\varphi)+\Pi^\text{soft,res}_\text{R}(k^2), \label{appeq:3dcorr}
\end{align}
where, by definition, the self-energy correction coincides with the resummed soft piece of the full 4d correlator. Writing $M^2_3=M^2_\phi+\delta M^2_3$ and equating Eqs.~\eqref{appeq:4dcorr} and \eqref{appeq:3dcorr} to $O(\lambda T^2)$ for soft momenta results in
\begin{align}
    \delta M^2_3(\varphi)=\Delta\Pi^\text{hard}_\text{R}(0)-M^2_\phi(\varphi)\Delta\Pi^{'\text{hard}}_\text{R}(0), \label{appeq:deltaM3}
\end{align}
where a diagrammatic expansion yields
\begin{equation}
\begin{split}
\label{appeq:massparameter}
    \Delta\Pi^{\text{hard}}_\text{R}(0)&= -4y^2\widetilde{\mathcal{I}}+8M^2_\psi y^2\widetilde{\mathcal{I}}_2+\frac{\lambda}{2}\mathcal{I}' \\
    &-\frac{G^2}{2}\mathcal{I}'_2+\delta M^2_\phi, \\
    \Delta\Pi^{'\text{hard}}_\text{R}(0)&=-\frac{8}{3}y^2M^2_\psi\widetilde{\mathcal{I}}_3+2y^2\widetilde{\mathcal{I}}_2+\frac{G^2}{6}\mathcal{I}'_3+\delta Z_\phi.
\end{split}
\end{equation}
Here, the prime above indicates that the corresponding zero mode is removed, and non-trivial numerators in tadpole sum-integrals were eliminated using integration-by-parts identities \cite{Nishimura:2012ee}. The results for the resulting sum-integrals are collected in Appendix~\ref{app:details}, while the one-loop mass and field counterterms $\delta M^2_\phi$ and $\delta Z_\phi$ are collected in Appendix~\ref{sec:betas}.

\subsection{Effective cubic coupling}
\label{app:effectivecubiccoupling}

Expressing the effective cubic coupling as $G_3=G+\delta G_3$, the one-loop correction to $\delta G_3$ can be obtained upon matching (renormalized) scalar three-point functions $\Gamma^{(3)}(p^0_i,\mathbf{p}_i)$ at vanishing external momenta in the static limit, i.e.
\begin{align}
    \Gamma^{(3)}_\text{4d}(p^0_i=0,\mathbf{p}_i\to 0)= \Gamma^{(3)}_\text{3d}(p^0_i=0,\mathbf{p}_i\to0),
\end{align}
where
\begin{align}
    \Gamma^{(3)}_\text{4d}(p^0_i=0,\mathbf{p}_i\to 0)&=G+\text{(4d loops)}, \\
    \Gamma^{(3)}_\text{3d}(p^0_i=0,\mathbf{p}_i\to 0)&=G_3+\text{(3d loops)}.   
\end{align}
For our three-loop example in Sec.~\ref{sec:threeloop}, the relevant contribution to $\delta G_3$ is extracted from the one-loop fermion loop diagram,
\begin{align}
    \delta G_3 \supset -2y^3\SumInt_{\widetilde{Q}} \frac{\text{tr}(-i\slashed{Q}+M_\psi)^3}{(Q^2+M_\psi^2)^3}.
\end{align}
The $y^3 M_\psi$ piece yields our required thermal counterterm entering Eq.~\eqref{threeloopexamplecounterterm}, leading to
\begin{align}
    \delta G_3\rvert_{\mathcal{M}}=24 y^3 M_\psi \SumInt_{\widetilde{Q}} \frac{1}{(Q^2+M^2_\psi)^2}.
\end{align}
Note that, to maintain clarity in the infrared cancellation example of Sec.~\ref{sec:threeloop}, here we retain the expressions in their bare form.

\section{\texorpdfstring{Two-loop example of \texttt{hotLTD} algorithm}{Two-loop example of hotLTD algorithm}}
\label{app:hotLTD-example}
In this Appendix, we present a worked-out example of the \texttt{hotLTD} algorithm \cite{Capatti} applied to the two-loop bosonic sunset integral $\mathcal{S}(m)$ defined in \cref{app:bosonicsunset}. This example illustrates the local subtraction of all UV and IR divergences and the analytic evaluation of Matsubara sums, leaving a finite expression suitable for numerical Monte Carlo integration over the remaining spatial momenta in $d=3$ dimensions. We denote
\begin{equation}
    \mathcal{S}(m) = \raisebox{-0.44\height}{\includegraphics[height=1.4cm]{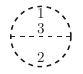}} = \SumInt_{P_1P_2} \prod_{i=1}^3\frac{1}{(Q_i^2+m^2)},
\end{equation}
with the propagator momenta $Q_1=P_1$, $Q_2=P_2$, and $Q_3=P_1-P_2$. We also denote the fully static (zero-mode) sunset integral as
\begin{equation}
    S_{3d}(m) = \raisebox{-0.44\height}{\includegraphics[height=1.4cm]{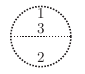}} = T^2\int_{\mathbf{p}_1\mathbf{p}_2} \prod_{i=1}^3\frac{1}{(\mathbf{q}_i^2+m^2)},
\end{equation}
which serves as an IR counterterm for $\mathcal{S}(m)$, canceling its logarithmic IR divergence in the limit $m\to 0$.

Although the combination $\mathcal{S}(m)-S_{3d}(m)$ is IR finite, it still contains UV divergences that must be subtracted at the integrand level before numerical integration. This is achieved using Bogoliubov's $R$-operation \cite{Bogoliubov:1957gp,Hepp:1966eg,Zimmermann:1969jj}, which acts recursively on a Feynman graph, generating counterterms involving the UV renormalization operator $K$. The operator $K$ isolates the superficial UV divergence of a subgraph and, in our thermal context, is applied only to its vacuum part, exploiting the exponential decay of thermal corrections in the UV. We use the vacuum definition of $K$ as detailed in \cite{Capatti:2022tit}.

Applying the $R$-operation to the IR-subtracted sunset integral with its corresponding vacuum contribution removed yields
\begin{widetext}
\begin{equation}
\begin{split}
R&\left( \raisebox{-0.44\height}{\includegraphics[height=1.4cm]{fig/R_op1_scalar.pdf}} - \raisebox{-0.44\height}{\includegraphics[height=1.4cm]{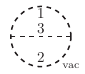}} - \raisebox{-0.44\height}{\includegraphics[height=1.4cm]{fig/R_op8_scalar.pdf}} \right) = \raisebox{-0.44\height}{\includegraphics[height=1.4cm]{fig/R_op1_scalar.pdf}} - \raisebox{-0.44\height}{\includegraphics[height=1.4cm]{fig/R_op1_scalar_vac.pdf}} - \raisebox{-0.44\height}{\includegraphics[height=1.4cm]{fig/R_op8_scalar.pdf}} - \raisebox{-0.44\height}{\includegraphics[height=1.4cm]{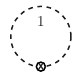}} * K\left( \raisebox{-0.44\height}{\includegraphics[height=1.4cm]{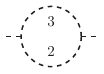}} \right) \\
&- \raisebox{-0.44\height}{\includegraphics[height=1.4cm]{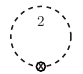}} * K\left( \raisebox{-0.44\height}{\includegraphics[height=1.4cm]{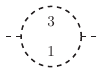}} \right) - \raisebox{-0.44\height}{\includegraphics[height=1.4cm]{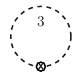}} * K\left( \raisebox{-0.44\height}{\includegraphics[height=1.4cm]{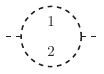}} \right) +  \raisebox{-0.44\height}{\includegraphics[height=1.4cm]{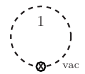}} * K\left( \raisebox{-0.44\height}{\includegraphics[height=1.4cm]{fig/R_op5_scalar.pdf}} \right) \\
&+ \raisebox{-0.44\height}{\includegraphics[height=1.4cm]{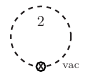}} * K\left( \raisebox{-0.44\height}{\includegraphics[height=1.4cm]{fig/R_op6_scalar.pdf}} \right) +  \raisebox{-0.44\height}{\includegraphics[height=1.4cm]{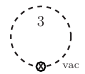}} * K\left( \raisebox{-0.44\height}{\includegraphics[height=1.4cm]{fig/R_op7_scalar.pdf}} \right) + K\left( \raisebox{-0.44\height}{\includegraphics[height=1.4cm]{fig/R_op8_scalar.pdf}} \right),\label{eq:Rop}
\end{split}
\end{equation}
\end{widetext}
where the final term subtracts the logarithmic UV divergence of the IR counterterm. The operator $K$ acts by extracting the superficial divergence of a subgraph's vacuum part through a Taylor expansion of the integrand in the limit where its loop momenta go to infinity. Specifically, each propagator is transformed by rescaling the loop momenta $P\to\epsilon^{-1}P$, yielding the following expression
\begin{equation}
\begin{split}
    &\frac{1}{(P+R)^2+m^2} \to \\
    & \frac{1}{\epsilon^{-2}(P^2+m_\mathrm{UV}^2)+2\epsilon^{-1} P\cdot R+R^2+m^2-m_\mathrm{UV}^2},
\end{split}
\end{equation}
where $R$ denotes an external momentum and $m_\mathrm{UV}$ is an arbitrary auxiliary mass introduced to regulate potential IR divergences in the counterterms. The integrand is then expanded around $\epsilon=0$ to the order dictated by the subgraph’s superficial degree of divergence.

As an explicit example, the action of $K$ on a logarithmically divergent self-energy subgraph in \cref{eq:Rop} gives
\begin{equation}
\begin{split}
K&\left(\raisebox{-0.42\height}{\includegraphics[height=1.4cm]{fig/R_op5_scalar.pdf}}\right) \\
&= K\left(\SumInt_{P_2}\frac{1}{(P_2^2+m^2)[(P_1-P_2)^2+m^2]}\right) \\
&= \int_{P_2}\frac{1}{\left(P_2^2+m_\mathrm{UV}^2\right)^2}, \label{eq:Kexp}
\end{split}
\end{equation}
where the expansion is truncated at leading order, as the subgraph has a logarithmic divergence (degree zero). To construct the corresponding full counterterm, we multiply the result by the reduced graph obtained by shrinking the UV subgraph to a point:
\begin{equation}
\begin{split}
&\raisebox{-0.44\height}{\includegraphics[height=1.4cm]{fig/R_op2_scalar.pdf}} * K\left(\raisebox{-0.44\height}{\includegraphics[height=1.4cm]{fig/R_op5_scalar.pdf}}\right) \\
&\qquad= \SumInt_{P_1}\frac{1}{P_1^2+m^2}\int_{P_2}\frac{1}{\left(P_2^2+m_\mathrm{UV}^2\right)^2}. \label{eq:Kexp2}
\end{split}
\end{equation}
The remaining counterterms in \cref{eq:Rop} are constructed analogously.

Before performing the numerical integration of the locally finite expression in \cref{eq:Rop}, we analytically evaluate all Matsubara sums and energy integrals in the spirit of Loop-Tree Duality. This procedure yields a three-dimensional integrand that is free of divergences and ready for numerical evaluation. Applying the algorithm presented in \cite{Capatti} (or equivalently through a straightforward calculation), we obtain
\begin{widetext}
\begin{equation}
\begin{split}
    &R\left( \raisebox{-0.44\height}{\includegraphics[height=1.4cm]{fig/R_op1_scalar.pdf}} - \raisebox{-0.44\height}{\includegraphics[height=1.4cm]{fig/R_op1_scalar_vac.pdf}} - \raisebox{-0.44\height}{\includegraphics[height=1.4cm]{fig/R_op8_scalar.pdf}} \right) \\
    &= \int_{\mathbf{p}_1\mathbf{p}_2} \sum_{\rho\in\{\pm 1\}^3}\Biggl\{ \frac{N^+(\rho_1E_1)N^-(\rho_2E_2)N^-(\rho_3E_3)-N^-(\rho_1E_1)N^+(\rho_2E_2)N^+(\rho_3E_3)}{2\rho_1E_12\rho_2E_22\rho_3E_3(\rho_1E_1-\rho_2E_2-\rho_3E_3)} \\
    &\quad- \frac{\Theta(\rho_1)\Theta(-\rho_2)\Theta(-\rho_3)+\Theta(-\rho_1)\Theta(\rho_2)\Theta(\rho_3)}{2\rho_1E_12\rho_2E_22\rho_3E_3(\rho_1E_1-\rho_2E_2-\rho_3E_3)} -\frac{T^2}{2E_1^22E_2^22E_3^2} \\
    &\quad-\frac{N^+(\rho_1E_1)\Theta(\rho_2)}{2\rho_1E_1(2\rho_2\tilde{E}_3)^3} - \frac{N^+(\rho_2E_2)\Theta(\rho_1)}{2\rho_2E_2(2\rho_1\tilde{E}_3)^3} - \frac{N^+(\rho_3E_3)\Theta(\rho_1)}{2\rho_3E_3(2\rho_1\tilde{E}_2)^3} \\
    &\quad+\frac{\Theta(\rho_1)\Theta(\rho_2)}{2\rho_1E_1(2\rho_2\tilde{E}_3)^3} + \frac{\Theta(\rho_2)\Theta(\rho_1)}{2\rho_2E_2(2\rho_1\tilde{E}_3)^3} + \frac{\Theta(\rho_3)\Theta(\rho_1)}{2\rho_3E_3(2\rho_1\tilde{E}_2)^3} + \frac{T^2}{2\tilde{E}_1^22\tilde{E}_2^22\tilde{E}_3^2} \Biggr\},\label{eq:subtractedsunset3d}
\end{split}
\end{equation}
\end{widetext}
where the thermal distributions are encoded in $N^\pm(E)\equiv[\pm1+\coth\left(\frac{E}{2T}\right)]/2$, and the on-shell energies are defined as $E_1=\sqrt{\mathbf{p}_1^2+m^2}$, $E_2=\sqrt{\mathbf{p}_2^2+m^2}$, and $E_3=\sqrt{(\mathbf{p}_1-\mathbf{p}_2)^2+m^2}$. The quantities $\tilde{E}_i$ denote the same energies evaluated at the auxiliary UV mass: $\tilde{E}_i=E_i\bigl|_{m\to m_\mathrm{UV}}$.

The expression in \cref{eq:subtractedsunset3d} can now be numerically integrated in $d=3$ spatial dimensions using spherical coordinates for each loop momentum. Variance reduction can be achieved through the multichanneling technique described in \cite{Capatti:2019edf,Navarrete:2024zgz}, in combination with adaptive importance sampling using standard Monte Carlo algorithms such as \texttt{Vegas} \cite{vegasgit}. Evaluating the integrand for different values of the ratio $m/T$ yields the function $b_T(q)$ introduced in \cref{app:sunsetresults}. Finally, the counterterms are computed analytically in dimensional regularization with $d=3-2\varepsilon$, and combined with the numerical result for the subtracted integral to reconstruct $\mathcal{S}(m)$, as presented in \cref{app:sunsetresults}.

The computation of the fermionic sunset integral $\tilde{\mathcal{S}}(m_1,m_2)$, defined in \cref{app:fermionicsunset}, closely parallels that of its bosonic counterpart, with the distinction that it is free of IR divergences and thus requires no associated counterterms. Similarly, the three-loop computation presented in \cref{sec:threeloop} adheres to the same overall algorithm; only the UV subtraction and Matsubara summation become more involved, but these steps are fully automated using the \texttt{hotLTD} algorithm \cite{Capatti}.

\section{Counterterms and renormalization group running}
\label{sec:betas}

In this Appendix, we collect all the required UV counterterms and $\beta$-functions relevant for the renormalization group running in our two-loop computation. They read:
\begin{widetext}    
\begin{align}
\delta \sigma &= \frac{1}{(4\pi)^2 \varepsilon} \Big( \frac{1}{2} g m^2_\phi - 4 y m^3_\psi \Big) \nonumber +\frac{1}{(4\pi)^4} \bigg \{ \frac{1}{\varepsilon^2} \Big[ g^3 + 8 y^2 g (m^2_\phi - 3 m^2_\psi) 
- 48 y^3 m_\psi (m^2_\phi + 2 m^2_\psi) + 4 \lambda g m^2_\phi \Big] \nonumber \\
&+ \frac{1}{\varepsilon} \Big[ -g^3 - 8 y^2 g (m^2_\phi - m^2_\psi) 
+ 16 y^3 m_\psi (m^2_\phi + 8 m^2_\psi) - 2 \lambda g m^2_\phi \Big] \bigg \},  
\end{align}

\begin{align}
\delta m^2_\phi &= \frac{1}{(4\pi)^2 \varepsilon} \frac{1}{2} \Big( g^2+ \lambda m^2_\phi - 24 y^2 m^2_\psi \Big) \nonumber +\frac{1}{(4\pi)^4} \bigg \{ \frac{1}{\varepsilon^2} \Big[ -96 y^3 g m_\psi -48 y^4 (m^2_\phi + 6 m^2_\psi) \nonumber \\
&+ 8 y^2 \lambda (m^2_\phi -3 m^2_\psi) +4 \lambda^2 m^2_\phi + g^2(8y^2+7 \lambda) \Big] \nonumber \\
&+ \frac{1}{\varepsilon} \Big[ 32 y^3 g m_\psi + 16 y^4 (m^2_\phi + 24 m^2_\psi) \nonumber  -8 y^2 \lambda (m^2_\phi - m^2_\psi) - 2 \lambda^2 m^2_\phi - g^2(8 y^2 + 5 \lambda) \Big] \bigg \}, 
\end{align}

\begin{align}
\delta g &= \frac{1}{(4\pi)^2 \varepsilon} \Big( \frac{3}{2} \lambda g - 24 y^3 m_\psi \Big) \nonumber + \frac{1}{(4\pi)^4} \bigg \{ \frac{6}{\varepsilon^2} \Big[ - 24 y^4 (g + 4y m_\psi) \nonumber + 4 y^2 \lambda (g-6y m_\psi) + 3 \lambda g \Big] \nonumber \\ 
&+ \frac{1}{\varepsilon} \Big[ 48 y^4 (g + 16 y m_\psi) - 24 y^2 \lambda (g - 2y m_\psi) - 12 \lambda^2 g \Big] \bigg \}, 
\end{align}

\begin{align}
\delta \lambda &= \frac{1}{(4\pi)^2 \varepsilon} \Big( \frac{3}{2} \lambda^2 - 24 y^4  \Big) \nonumber +\frac{1}{(4\pi)^4} \bigg [ -\frac{6}{\varepsilon^2} \Big( 96 y^6 + 48 y^4 \lambda - 4 y^2 \lambda^2 - 3 \lambda^3 \Big) \nonumber \\
&+ \frac{12}{\varepsilon} (4y^2-\lambda) \Big( 16 y^4 + 6 y^2 \lambda + \lambda^2 \Big) \bigg ],
\end{align}

\begin{align}
\delta Z_\phi = \frac{1}{(4\pi)^2 \varepsilon } \Big( -2 y^2 \Big) \nonumber +\frac{1}{(4\pi)^4} \bigg[ y^4\Big( -\frac{3}{\varepsilon^2} + \frac{5}{2\varepsilon} \Big)  + \frac{\lambda^2}{24} \frac{1}{\varepsilon} \bigg], 
\end{align}
\begin{align}
\delta m_\psi &= \frac{1}{(4\pi)^2 \varepsilon} \Big( y^2 m_\psi \Big) ,  
\end{align}
\begin{align}
\delta y &= \frac{1}{(4\pi)^2 \varepsilon} \Big( y^3 \Big) , 
\end{align}
\begin{align}
\delta Z_\psi &= \frac{1}{(4\pi)^2 \varepsilon} \Big( -\frac{1}{2} y^2 \Big) , 
\end{align}
and
\begin{align}
\overline{\Lambda} \frac{d}{d \overline{\Lambda}} \sigma &= \frac{1}{(4\pi)^2} \Big( -8 y m^3_\psi + 2 y^2 \sigma + g m^2_\phi \Big) \nonumber + \frac{1}{(4\pi)^4} \Big[ 4g y^2 (m^2_\psi - m^2_\phi) + 8 y^3 m^2_\phi m_\psi \nonumber \\
&  + 64 y^3 m^3_\psi - \lambda g m^2_\phi - \sigma (5 y^4 + \frac{\lambda^2}{12}) -\frac{1}{2}g^3 \Big], 
\end{align}

\begin{align}
\overline{\Lambda} \frac{d}{d\overline{\Lambda}} m^2_\phi &= \frac{1}{(4\pi)^2} \Big[ g^2 + m^2_\phi(\lambda + 4 y^2) -24 y^2 m^2_\psi \Big] \nonumber + \frac{1}{(4\pi)^4} \Big[  16 y^3 g m_\psi - 2 y^4 (m^2_\phi - 96 m^2_\psi) \nonumber \\
& + 4 y^2 \lambda (m^2_\psi - m^2_\phi) - \frac{7}{6} \lambda^2 m^2_\phi - g^2(4y^2 + \frac{5}{2}\lambda) \Big], 
\end{align}

\begin{align}
\overline{\Lambda} \frac{d}{d \overline{\Lambda}} g &= \frac{1}{(4\pi)^2} \Big[ 3g( \lambda  + 2 y^2) - 48 y^3 m_\psi \Big] \nonumber + \frac{1}{(4\pi)^4} \Big[ 24 y^3 m_\psi (16 y^2 + \lambda) \nonumber  + g(9 y^4 - 12 y^2 \lambda -\frac{25}{4} \lambda^2) \Big], 
\end{align}

\begin{align}
\label{eq:lam-run}
\overline{\Lambda} \frac{d}{d \overline{\Lambda}} \lambda &= \frac{1}{(4\pi)^2} \Big( 3 \lambda^2 + 8 y^2 \lambda - 48 y^4 \Big) \nonumber + \frac{1}{(4\pi)^4} \Big( 384 y^6 + 28 y^4 \lambda - 12 y^2 \lambda^2 -\frac{19}{3} \lambda^3 \Big), 
\end{align}

\begin{align}
\overline{\Lambda} \frac{d}{d \overline{\Lambda}} m_\psi &= \frac{1}{(4\pi)^2} \Big( 3 y^2 m_\psi \Big), 
\end{align}

\begin{align}
\overline{\Lambda} \frac{d}{d \overline{\Lambda}} y &= \frac{1}{(4\pi)^2} \Big( 5 y^3 \Big).
\end{align}
\end{widetext}
For scalar parameters that appear at tree-level in the effective potential, we have included results up to and including two loops, while counterterms and running for $m_\psi$ and $y$ are only required to one-loop order.
Note that the running of $\lambda$ in Eq.~\eqref{eq:lam-run} includes a large negative contribution $-48 y^4$ and it is this term that can cause issues with the zero-temperature stability, as we have discussed in Sec.~\ref{sec:thermo}.

Together with the anomalous dimension
\begin{align}
\gamma &= \frac{1}{(4\pi)^2} \Big( -2 y^2 \Big) + \frac{1}{(4\pi)^4} \Big( 5 y^4 + \frac{1}{12} \lambda^2 \Big),   
\end{align}
which describes the running of the field $\gamma \equiv \frac{1}{\varphi} \overline{\Lambda} \frac{d}{d \overline{\Lambda}} \varphi$,
the beta functions satisfy the Callan-Symanzik equation for the zero temperature effective potential, resulting from
$\overline{\Lambda} \frac{d}{d \overline{\Lambda}} V^{T=0}_{\text{eff}}(\varphi) = 0$, 
which holds order by order. Technically, running of the tree-level potential cancels explicit logarithms of the renormalization scale at one- and two-loop orders, while one-loop running inside the one-loop potential cancels explicit double logarithms at two loops.

\section{\texorpdfstring{Thermodynamics using 3d EFT in a high-$T$ expansion}{Thermodynamics using 3d EFT in a high-T expansion}}
\label{sec:3D-EFT}

In this Appendix, we provide the details of the 3d EFT–based computation of thermodynamic quantities within the high-$T$ expansion, as used in Sec.~\ref{sec:thermo}. For this purpose, we have followed and reproduced the results of Refs.~\cite{Gould:2021dzl,Gould:2023jbz}.

The effective parameters of the 3d EFT can be directly obtained by expanding Eqs.~\eqref{eq:naive1} and \eqref{eq:naive2} in the high-$T$ limit, using the results of Sec.~\ref{sec:pop}. This is justified because the effective potential serves as the generator of all required static Green’s functions for the matching procedure \cite{Kajantie:1995dw}. Only hard-mode contributions enter the matching relations, which determine the parameters of the 3d EFT. These are combined with the field normalization between the 3d and 4d theories, computed separately from the quadratic external momentum part of the two-point function:
\begin{align}
\phi^2_{\text{3d}} &= \frac{\phi^2_{\text{4d}}}{T}\Big( 1 + \frac{1}{(4\pi)^2} 2y^2 L_f \Big).
\end{align}
The matching relations up to and including NLO then read: 
\begin{widetext}
\begin{align}
\label{eq:sigma3}
\sigma_3(\overline{\Lambda}_\text{3d}) &= \frac{1}{\sqrt{T}} \bigg\{ \Big [ \sigma + \frac{T^2}{24}\left (g + 4 y m_\psi \right ) \Big ]  \Big( 1 - \frac{y^2L_f}{(4\pi)^2} \Big) + \frac{1}{(4\pi)^2} \bigg[ -\frac{1}{2} L_b g m^2_\phi + 4 L_f y m^3_\psi  \\
&+ T^2 \Big[ -\frac{L_b \lambda g}{16}  + \Big( \frac{L_f}{2} - \frac{4}{3} \ln(2) \Big) y^3 m_\psi \nonumber - \Big( \frac{L_f}{12} - \frac{1}{2} \ln(2) \Big) y^2 g \Big] -\frac{\lambda_3 g_3}{6}  \Big( c + \ln \frac{3 T}{\overline{\Lambda}_\text{3d}} \Big) \bigg]  \bigg\},
\end{align}

\begin{align}
\label{eq:mphisq3}
m^2_3(\overline{\Lambda}_\text{3d}) &= \bigg [ m^2_\phi + \frac{T^2}{24} \Big( \lambda + 2y^2 \Big) \bigg ] \nonumber \Big( 1 - \frac{ 2 y^2 L_f}{(4\pi)^2} \Big) + \frac{1}{(4\pi)^2} \bigg \{ - \frac{1}{2} L_b g^2  + 12 L_f y^2 m^2_\psi - \frac{1}{2} \lambda m^2_\phi \\
& + T^2 \Big[ -\frac{L_b \lambda^2}{16}  + \Big( \frac{L_f}{2} - \frac{4}{3} \ln(3) \Big) y^4 - \Big( \frac{L_f}{2} - \frac{1}{2} \ln(2) \Big) y^2 \lambda \Big]  - \frac{\lambda^2_3}{6}  \Big( c + \ln \frac{3 T}{\overline{\Lambda}_\text{3d}} \Big)  \bigg \}, 
\end{align}

\begin{align}
\label{eq:g3}
g_3 &= \sqrt{T} \bigg [ g + \frac{1}{(4\pi)^2}\Big( -\frac32 g \lambda L_b + 24 y^3 m_\psi L_f - 3 y^2 g L_f \Big) \bigg ], 
\end{align}

\begin{align}
\label{eq:lam3}
\lambda_3 &= T \bigg [ \lambda + \frac{1}{(4\pi)^2} \Big( -\frac32 \lambda^2 L_b \nonumber + 24 y^4 L_f - 4 \lambda y^2 L_f \Big) \bigg ].
\end{align}
Here, the 4d parameters depend on $\Lbar$ and we used the following shorthand notation
\begin{align}
L_b &\equiv 2\ln  \frac{\overline{\Lambda}}{4\pi e^{-\gamma}T}  , \\
L_f &\equiv L_b(\overline{\Lambda}) + 4 \ln(2), \\
c &\equiv \frac{1}{2} \Big( \ln \frac{8\pi}{9} + \frac{\zeta_2'}{\zeta_2} - 2 \gamma \Big),
\end{align}
\end{widetext}
where $\zeta$ is the Riemann $\zeta$-function.

All 3d parameters are independent of the 4d renormalization scale $\overline{\Lambda}$ at NLO, which becomes evident when one-loop running is applied to the leading-order terms. Concretely, the LO relations correspond to the parts of the above matching expressions that do not involve $1/(4\pi)^2$ factors -- i.e., one-loop terms for the tadpole and thermal mass, and tree-level terms for the couplings. In particular, the running of parameters within the one-loop thermal mass and tadpole cancels the explicit $T^2$-proportional logarithmic terms that appear at two-loop order \cite{Braaten:1995cm,Gould:2021oba}. Additional NLO contributions arise at one-loop, especially from mass corrections in the high-$T$ expansion, as seen in Eqs.~\eqref{eq:sigma3} and \eqref{eq:mphisq3}. The tadpole and thermal mass depend on the 3d renormalization scale $\overline{\Lambda}_\text{3d}$, whereas the couplings $g_3$ and $\lambda_3$ are renormalization group invariant in the 3d theory \cite{Farakos:1994kx}.

We further emphasize that the logarithmic $y^4 L_f$ term contributing to the thermal self-coupling $\lambda_3$ in Eq.~\eqref{eq:lam3} highlights the challenges in defining the $\overline{\text{MS}}$ parameters, as discussed in Sec.~\ref{sec:thermo}: this term precisely cancels the corresponding running in Eq.~\eqref{eq:lam-run}. However, in the presence of a large hierarchy $\lambda \ll y$, the fermionic contributions can run $\lambda$ negative, making the model unstable.

As explained in  Refs.~\cite{Gould:2021dzl,Gould:2023jbz}, within the 3d EFT the critical temperature is determined from the condition
\begin{align}
\tilde{\sigma}_3 \equiv \sigma_3 + \frac{g^3_3}{3 \lambda^2_3} -\frac{g_3}{\lambda_3} m^2_{\phi,3}
\end{align}
with $\tilde{\sigma}_3(T_c) = 0$. This condition is exact within the 3d EFT and does not depend on higher-order corrections within the EFT itself, but only on corrections to the matching parameters. Using the LO matching relations yields a simple analytic expression for the critical temperature:
\begin{align}
\label{eq:TcLO}
T^{\text{LO}}_{c} &= \frac{\sqrt{2 g^3 - 6 \lambda g m^2_\phi + 6 \lambda^2 \sigma}}{\sqrt{y \lambda (y g - \lambda m_\psi)}},
\end{align}
which corresponds to the LO curve shown in Fig.~\ref{fig:results}. At NLO, however, a closed-form expression for $T_c$ is no longer available, but the condition $\tilde{\sigma}_3(T_c) = 0$ can still be solved numerically. For large values of $y$ in Fig.~\ref{fig:results}, $\tilde{\sigma}_3$ develops a lower bound as temperature decreases, and eventually its minimum becomes positive, implying that no solution for $T_c$ exists at NLO. In terms of the effective potential, this reflects the fact that for large $y$, when transitions cease to exist, the value of the effective potential in the broken minimum initially decreases as temperature drops, but then begins to rise again, never becoming degenerate with the high-$T$ minimum near the origin. This peculiar feature clearly signals the breakdown of the high-$T$ expansion in the matching relations, as it is entirely absent in the full computation performed without relying on the high-$T$ expansion.

Finally, we compute the latent heat in a strict expansion akin to Refs.~\cite{Gould:2021dzl,Gould:2023jbz}, that is 
\begin{align}
L(T_c) = T_c \eta(\tilde{\sigma}_3) \Delta \langle \phi_3 \rangle, 
\end{align}
where the eta-function $\eta(\tilde{\sigma}_3) \equiv T d\tilde{\sigma}_3/dT$ at $T_c$ \cite{Gould:2019qek} and the jump in the scalar condensate \cite{Farakos:1994xh} admits a strict expansion
\begin{align}
\label{eq:condensate}
\langle \phi_3 \rangle &= v_0 \bigg[ 2 + \sqrt{3} \alpha_3 +  \bigg( \frac{1}{2} + \ln  \frac{\overline{\Lambda}_\text{3d}}{\sqrt{3 \lambda_3} v_0}  \bigg) \alpha^2_3 \nonumber \\
&-1.15232 \alpha^3_3  + O(\alpha^4_3) \bigg].
\end{align}
Here, we use the shorthand notation $v_0 \equiv \sqrt{-6 \tilde{m}^2_{\phi,3}/\lambda_3}$ for the order parameter with $\tilde{m}^2_{\phi,3} = m^2_{\phi,3} - g^2_3/(2\lambda_3)$ and the effective expansion parameter $\alpha_3 \equiv \sqrt{\lambda_3}/(4\pi|v_0|)$. The numerical factor in the term $O(\alpha^3_3)$ in Eq.~\eqref{eq:condensate} arises from a numerical integration that results from the three-loop calculation of the effective potential \cite{Rajantie:1996np}.

When $y \in [0.3,1.0]$, the expansion parameter $\alpha_3 \in [0.8,0.05]$ decreases with increasing $y$. As a result, the strict expansion of Eq.~\eqref{eq:condensate} converges rapidly for large $y$, making higher-order corrections within the EFT negligible. However, as discussed above, the EFT construction itself becomes unreliable in this regime. This is because $\alpha_3$ is inversely proportional to the order parameter $v_0$: for stronger transitions -- characterized by a larger jump in the order parameter -- perturbation theory converges more quickly. At the same time, a strong transition induces a large mass for the field driving it (in our case, the fermion), which invalidates the assumptions underlying the high-$T$ expansion and thus the EFT description. We emphasize that these are generic features of first-order phase transitions, observed also in gauge-Higgs theories; see Ref.~\cite{Ekstedt:2024etx} and Appendix~\ref{sec:portal-sector} for further discussion.

Analogously to Eq.~\eqref{eq:TcLO}, the LO result for the latent heat can also be expressed in closed form. However, we omit the explicit expression here due to its slightly greater length. We note that at NLO, both $T_c$ and $L$ are independent of the 3d renormalization scale $\overline{\Lambda}_\text{3d}$. This follows from the fact that $\tilde{\sigma}_3$ is entirely independent of $\overline{\Lambda}_\text{3d}$, and the condensate, when computed via the strict expansion, is likewise independent of $\overline{\Lambda}_\text{3d}$ up to the order considered. This is because the running of the thermal mass in the leading-order term cancels the explicit logarithmic dependence that appears at two-loop order in the $O(\alpha_3^2)$ term. Moreover, since the one-loop contribution at $O(\alpha_3)$ is independent of the thermal mass, there is no logarithmic dependence on the 3d scale at three-loop order in the $O(\alpha_3^3)$ term.

\section{A brief excursion into Higgs portal models} 
\label{sec:portal-sector}

In this Appendix, we further motivate why it is important to extend our newly developed framework to strong transitions in many BSM theories --- that is, why we can anticipate a breakdown of the high-$T$ expansion for these strong EW phase transitions. To illustrate this, let us consider the SM augmented by a new scalar field, $\Phi$. For concreteness, we assume here that $\Phi$ is an inert doublet, as studied in e.g. Refs.~\cite{Blinov:2015vma,Laine:2017hdk,Kainulainen:2019kyp}, but emphasize that the following discussion is quite general and applies to other similar extensions as well, regardless of the representation of $\Phi$ under the SM gauge groups. Hence, while keeping the discussion general, we will concretely exemplify the Inert Doublet Model (IDM) where appropriate.

In the IDM, there are three portal interaction couplings, $\lambda_3, \lambda_4,$ and $\lambda_5$, which couple the new doublet to the SM Higgs field. The doublets do not mix, and only the SM doublet acquires a VEV. For simplicity, we assume $\lambda_4 = \lambda_5$ and both are negative, so that the combination $\lambda_3 + \lambda_4 + \lambda_5$ is small. Under these assumptions, it is essentially $\lambda_3$ that controls the portal interaction with the Higgs, and below we denote this portal coupling as $\lambda_{\text{p}}$ for generality.

For $\Phi$ to trigger a strong first-order, one-step electroweak phase transition, we assume it to be heavy at high temperatures (though not necessarily at low temperatures) compared to the Higgs scalar undergoing the transition. This allows $\Phi$ to contribute to the thermal effective potential in the Higgs phase at leading order, when integrated out to compute the Higgs phase free energy \cite{Ekstedt:2022zro,Gould:2023ovu}. In such an approach, $\Phi$ induces a large cubic term in the leading order thermal effective potential, which enhances the potential barrier between the minima. The strength of the transition is approximately inversely proportional to the square of the ratio $\lambda/\lambda_{\text{p}}$ of the Higgs self-coupling to the portal coupling \cite{Ekstedt:2024etx}.

To concretely demonstrate this, the leading order thermal effective potential at high temperatures \cite{Gould:2023ovu}, assuming that the high-$T$ expansion is valid, can be computed in the dimensionally reduced effective theory of the IDM \cite{Losada:1996ju,Andersen:1998br,Andersen:2017ika,Gorda:2018hvi}. In terms of the 3d Higgs background field $v_3$ the (approximate) result reads 
\begin{align}
\label{eq:VLO-IDM}
V^{\text{LO}}_{\text{3d}}(v_3) &=\frac{1}{2} m^2_{3} v^2_3 + \frac{1}{4} \lambda_{3} v^4_3 \nonumber \\
&- \frac{1}{12\pi} \bigg( 6\Big( \frac{1}{4} g^2_3 v^2_3  \Big)^{\frac{3}{2}} + 3\Big( m^2_D +  \frac{1}{2} h_3 v^2_3  \Big)^{\frac{3}{2}} \nonumber \\
&+ N_s \Big( \mu^2_3 +  \frac{1}{2} \lambda_{\text{p},3} v^2_3  \Big)^{\frac{3}{2}}\bigg), \nonumber  \\
&\approx  \eta^2_3 \bigg( \frac{1}{2} {y'}^2 \widetilde{v}^2_3 + \frac{1}{4} x' \widetilde{v}^4_3 - \frac{\widetilde{v}_3^3}{16\pi} \bigg),    
\end{align}
where $m^2_3$ and $\mu^2_3$ are the Higgs and $\Phi$ thermal masses, respectively, $g_3$ is the thermal gauge coupling, $m^2_D$ is the Debye mass of the SU(2) temporal gauge field component $A^a_0$ (for simplicity, we ignore contributions of U(1) and SU(3) sectors here) and $h_3$ is the coupling between $A^a_0$ and the Higgs field.
$N_s = 3$ is the number of heavy/soft scalar fields in the IDM, and we have utilized aforementioned assumptions for $\lambda_3$, $\lambda_4$ and $\lambda_5$.

To arrive to the last line in Eq.~\eqref{eq:VLO-IDM} we have assumed that $m^2_D$ and $\mu^2_3$ are dominated by the field-dependent contributions with portal couplings $h_3 \approx g^2_3/2 = g^2 T/2$ and $\lambda_{\text{p},3} \approx T \lambda_{\text{p}}$. This approximation describes very strong transitions that involve large value of the background field at the critical temperature, c.f. Ref.~\cite{Bernardo:2025vkz}. 
In the last line of Eq.~\eqref{eq:VLO-IDM} we also defined dimensionless forms of the thermal mass and the thermal self-coupling via
\begin{align}
\label{eq:xy}
y' &\equiv \frac{m^2_3}{\eta^{\frac{4}{3}}_3} , \quad \quad x' \equiv \frac{\lambda_3}{\eta^{\frac{2}{3}}_3}, 
\end{align}
through a scaling transformation $v_3 = \eta^{\frac{1}{3}}_3 \widetilde{v}_3$, where the field $\widetilde{v}_3$ is dimensionless and the effective cubic coupling can be approximated as 
\begin{align}
\eta_3 \simeq \mathcal{E} g^3_3 + \sqrt{2} \lambda^{\frac{3}{2}}_{\text{p},3} \underset{\lambda_{\text{p}} \gtrsim 1}{\approx} \sqrt{2} \lambda^{\frac{3}{2}}_{\text{p},3}.
\end{align}
Here the factor $\mathcal{E}=3/2$ arises from the enhancement due to temporal gauge field component \cite{Bernardo:2025vkz}, yet for $\lambda_{\text{p}} \gtrsim 1$ contributions of $\Phi$ dominate over the terms induced by the gauge field. 

In the pure SM, an analogous parameter to $x'$ is defined as $x \equiv \lambda_3/g^2_3 \approx \lambda/g^2$, which --- for the observed Higgs boson mass --- leads to $x_c \equiv x(T_c) \gg x_* \simeq 0.1$ indicating a crossover \cite{Kajantie:1995kf,Gould:2022ran}. In the IDM, the new heavy doublet can essentially decrease the Higgs thermal self-interaction coupling $x'$ provided that portal interaction coupling $\lambda_{\text{p}}$ is sufficiently large, and can hence be expected to trigger a strong first-order phase transition when $x'(T_c) \equiv x_c' \ll x_*$.   

The effective thermal mass $y'$ of Eq.~\eqref{eq:xy} controls when the critical temperature occurs, and at $y'(T_c) \equiv y_c'(x) = 1/(128\pi^2 x')$ minima of the potential are degenerate. Released latent heat can be computed as \cite{Farakos:1994xh}
\begin{align}
\frac{L}{T^4_c} \approx \frac{\eta^2_3}{T^3_c} \Big( T_c \frac{dy'}{dT}|_{T_c} \Big)  \Big(\frac{\partial}{\partial y'} \Delta \widetilde{V}^{\text{LO}}_{\text{3d}}(x',y')\Big)|_{y_c'}, 
\end{align}
where the difference of the quadratic scalar condensate $\Delta \langle \phi^\dagger\phi \rangle \equiv \Delta \partial/\partial y' [ \widetilde{V}^{\text{LO}}_{\text{3d}}(x',y')]|_{y_c'} = 1/(128\pi^2 {x'}^2)$. Hence, we observe that the transition strength is proportional to ${x'}^{-2} \approx 2^{\frac{2}{3}} \lambda^2_{\text{p}}/\lambda^2$, i.e. larger portal coupling leads to stronger transitions.

So far, we have assumed that the 3d EFT description, and hence the high-$T$ expansion, is valid. 
This requires that the thermal mass of the new doublet in the Higgs phase is suppressed compared to hard scale, i.e. that
\begin{align}
\frac{1}{2} \lambda_{\text{p},3} v^2_3 \ll (\pi T)^2,
\end{align}
which -- using the leading order relation $\widetilde{v}_{3,c} = 1/(8\pi x')$ at the critical temperature -- implies an upper bound  
\begin{align}
\label{eq:portal-upper-bound}
\lambda_{\text{p}} \ll \sqrt{8 \pi^2 \lambda} \approx 2.81.
\end{align}
This particular numerical value for the upper bound on the portal coupling depends on $N_s$ in Eq.~\eqref{eq:VLO-IDM}, and is hence different for other models, yet the rationale of its derivation is general.

Since transitions become stronger for increasing $\lambda_{\text{p}}$, the upper bound in Eq.~\eqref{eq:portal-upper-bound} indicates that the use of the 3d EFT becomes unreliable for the strongest transitions at large portal couplings in the IDM. Hence, the thermodynamics of this regime remain elusive to the conventional 3d EFT approach and require methods that do not rely on the high-$T$ expansion. This challenge has been addressed previously in Ref.~\cite{Laine:2017hdk} for the equilibrium thermodynamics of the IDM, and can be revisited with refined tools developed in this work at hand. Connecting these strongest transitions to gravitational wave predictions requires consistent and accurate determination of the bubble nucleation rate without high-$T$ expansion, and is so far uncharted territory.

\bibliography{ref}

\end{document}